\documentclass{pasj01}

\usepackage{lscape} 

\begin{document}
\SetRunningHead{Author(s) in page-head}{Running Head}

\title{Nature of bright C-complex asteroids}
\author{%
   Sunao \textsc{Hasegawa},\altaffilmark{1,*}
   Toshihiro \textsc{Kasuga},\altaffilmark{2}
   Fumihiko \textsc{Usui},\altaffilmark{3}
   and
   Daisuke \textsc{Kuroda}\altaffilmark{4}
}
 \altaffiltext{1}{Institute of Space and Astronautical Science, Japan Aerospace Exploration Agency, 3-1-1 Yoshinodai, Chuo-ku, Sagamihara 252-5210, Japan}
 \email{hasehase@isas.jaxa.jp}
 \altaffiltext{2}{Public Relations Center, National Astronomical Observatory of Japan, 2-21-1 Osawa, Mitaka-shi, Tokyo 181-8588, Japan}
 \altaffiltext{3}{Center for Planetary Science, Graduate School of Science, Kobe University, 7-1-48, Minatojima-minamimachi, Chuo-Ku, Kobe 650-0047, Japan}
 \altaffiltext{4}{Okayama Astronomical Observatory, Graduate School of Science, Kyoto University, 3037-5 Honjo, Kamogata-cho, Asakuchi, Okayama 719-0232, Japan}


\maketitle

\begin{abstract}
Most C-complex asteroids have albedo values less than 0.1, but there are some high-albedo (bright) C-complex asteroids with albedo values exceeding 0.1.
To reveal the nature and origin of bright C-complex asteroids, we conducted spectroscopic observations of the asteroids in visible and near-infrared wavelength regions.
As a result, the bright B-, C-, and Ch-type (Bus) asteroids, which are subclasses of the Bus C-complex, are classified as DeMeo C-type asteroids with concave curvature, B-, Xn-, and K-type asteroids.
Analogue meteorites and material (CV/CK chondrites, enstatite chondrites/achondrites, and salts) associated with these spectral types of asteroids are thought to be composed of minerals and material exposed to high temperatures. 
A comparison of the results obtained in this study with the SDSS photometric data suggests that salts may have occurred in the parent bodies of 24 Themis and 10 Hygiea, as well as 2 Pallas.
The bright C-complex asteroids in other C-complex families were likely caused by impact heating.
Bright C-complex asteroids that do not belong to any families are likely to be impact metamorphosed carbonaceous chondrites, CV/CK chondrites, or enstatite chondrites/achondrites.
\end{abstract}

\section{Introduction}
In the 1970s, within several years after the first spectrophotometric study of asteroids \citep{McCord1970}, asteroids with flat reflectance spectra over the visible wavelength range were discovered (\cite{Johnson1973}; \cite{McCord1974}; \cite{Chapman1975}).
Since these asteroids have spectral and physical properties similar to carbonaceous chondrites (e.g., \cite{Johnson1973}; \cite{McCord1974}; \cite{Chapman1975}) and interplanetary dust particles (e.g., \cite{Vernazza2015}; \cite{Hasegawa2017}), the asteroids are considered as parent bodies of these chondrites.
These asteroids with flat reflectance spectra over the visible wavelength range are defined or categorized as C-complex/group/type asteroids (hereafter C-complex asteroids) (\cite{Chapman1975}; \cite{Tholen1984}; \cite{Bus2002b}).
The albedo of C-complex asteroids is typically below 0.1 (\cite{Chapman1975}; \cite{Tedesco1989}).
The major trends in the distribution of the albedo of C-complex asteroids have not shown significant differences in recent studies that used infrared astronomical satellites dedicated to all-sky surveys, AKARI \citep{Murakami2007} and WISE \citep{Wright2010}; however, small C-complex asteroids with albedos greater than 0.1 were found  (\authorcite{Mainzer2011b} \yearcite{Mainzer2011b}, \yearcite{Mainzer2012}; \cite{Usui2013}).

To examine characteristics of C-complex asteroids with the geometric albedo of more than 0.1 (hereafter ``bright'' C-complex asteroids) in more detail, \authorcite{Kasuga2013} (\yearcite{Kasuga2013}, \yearcite{Kasuga2015}) conducted near-infrared spectroscopic observations of bright C-complex asteroids in the outer main-belt and no signs of water ice were found in the bright C-complex asteroids.
These studies argue that Mg-rich amorphous and/or crystalline silicates on the surface of the asteroids can be responsible for the high-albedo values.
The presence of crystalline silicate implies the occurrence of high-temperature environments during planetesimal formation, which was confirmed in some of the observed bright C-complex asteroids.

It is essential to classify the reflectance spectra of asteroids in order for knowing the properties of the surface materials on asteroids.
The Bus taxonomy \citep{Bus2002b} that classifies asteroids using visible reflectance spectra according to principal component analysis (PCA) and the Bus--DeMeo taxonomy (\cite{DeMeo2009}; \cite{Binzel2019}) does those using the visible and near-infrared spectra according to PCA are the most popular and important classification methods for classifying asteroids.
In particular, the Bus--DeMeo taxonomy is based on spectra with more wide wavelength range, so it is expected that more detailed asteroid properties will be extracted.

In this study, we made spectroscopic observations in visible and near-infrared wavelength regions, to obtain the reflectance spectra of the bright C-complex asteroids.
On the basis of obtained near-infrared spectroscopic results of bright C-complex asteroids, we discuss the origin of the bright C-complex asteroids using large amounts of the spectrophotomertic data of asteroids.
The subsequent sections of this paper describe the observations and data reduction procedures (section 2), presents the spectroscopic and taxonomic results of bight C-complex asteroids (section 3) and the boundaries of albedoes for bright C-complex asteroids (section 4), and discusses the nature and origin of the asteroids (section 5).

\section{Observations and data reduction procedures}
In order to carry out the Bus--DeMeo taxonomy of bright C-complex asteroids for which visible spectra have already been known in the literatures, near-infrared spectroscopic observations for the asteroids were performed in this study.
Main candidates for observations in near-infrared wavelength regions were assigned bright C-complex asteroids on the basis of the Bus taxonomy (\cite{Bus2002b}; \cite{Lazzaro2004})\footnotemark[1] listed in the AKARI asteroid catalogs (\cite{Usui2011}; \cite{Hasegawa2013}).
A bright T-type asteroid, which typically has low albedo characteristics, was also observed (979 Ilsewa).
Besides, to increase the sample of bright C-complex asteroids, visible spectroscopic observations for bright ``unknown-type'' asteroids were carried out.
Unknown spectral type candidates were selected from bright asteroids in the outer main-belt region, where many C-complex asteroids are distributed, listed in the AKARI asteroid catalogs.
Furthermore, spectroscopic observations of several ``known as'' C-complex asteroids with low albedo, which are located in the outer main-belt region, were also conducted to evaluate the reliability of the obtained reflectance spectra.
\footnotetext[1]{Taxonomic data in \citet{Carvano2010} were not available when selecting the candidates in this study.}

The spectroscopic observations for bright C-complex asteroids were conducted at two observatories: the Subaru Telescope, National Astronomical Observatory of Japan (NAOJ), National Institute of Natural Sciences (NINS) on Mt. Mauna Kea in Hawaii, USA (Minor Planet Center (MPC) code: 568) from 2012 February to 2013 April; the Okayama Astrophysical Observatory, NAOJ, NINS, in Okayama, Japan (MPC code 371) during December 2013 and January 2014. 
Ten and eleven spectroscopic observations of the asteroids were made with the Subaru telescope and Okayama Astrophysical Observatory, respectively.
The nightly observational details for the spectroscopy are listed in table \ref{tab:obs}.

\begin{landscape}
\setlength{\headsep}{60mm} 
\setlength{\textheight}{160mm} 
\setlength{\tabcolsep}{3pt} 
{
\footnotesize
\begin{longtable}{llllllllll}
  \caption{Spectroscopic circumstances of the asteroids.}\label{tab:obs}
  \hline
    Num & Name & Date (UT) & Start--End time  & $R_{\rm h}$\footnotemark[$*$] & $\Delta$\footnotemark[$*$] & $\alpha$\footnotemark[$*$] & Airmass & Instr. & Solar analogue\\
        & & [YY.MM.DD] & [hr:min] &[au] & [au] & [$\timeform{D}$] & &  &\\
\endfirsthead
  \hline
    Num & Name & Date (UT) & Start--End time  & $R_{\rm h}$\footnotemark[$*$] & $\Delta$\footnotemark[$*$] & $\alpha$\footnotemark[$*$] & Airmass & Instr. & Solar analogue\\
  \hline
\endhead
  \hline
\endfoot
  \hline
\multicolumn{1}{@{}l}{\rlap{\parbox[t]{1.0\textwidth}{\small
\footnotemark[$*$]The heliocentric distance ($R_{\rm h}$), geocentric distance ($\Delta$), and phase angle ($\alpha$) for asteroid observations were obtained from the NASA/ Jet Propulsion Laboratory (JPL) HORIZONS ephemeris generator system.\footnotemark[3]\\
}}}
\endlastfoot
  \hline
320  & Katharina  & 2013.01.06 & 08:54--09:49 & 2.927 & 1.994 & 7.4  & 1.01--1.02 & IRCS  & HD283798\\
366  & Vincentina & 2012.11.25 & 06:21--07:17 & 3.044 & 2.573 & 17.8 & 1.08--1.17 & IRCS  & SA115-271\\ 
419  & Aurelia    & 2012.02.17 & 12:01--12:25 & 2.709 & 1.761 & 7.3  & 1.07--1.08 & IRCS  & SA103-272, HIP52192\\
840  & Zenobia    & 2012.11.25 & 12:53--14:16 & 3.433 & 2.915 & 15.2 & 1.02--1.13 & IRCS  & HD79078, SA98-978\\
979  & Ilsewa     & 2012.11.25 & 04:55--05:58 & 2.736 & 2.626 & 23.2 & 1.12--1.25 & IRCS  & SA115-271\\ 
981  & Martina    & 2012.02.17 & 08:13--09:22 & 3.496 & 2.633 & 9.1  & 1.00--1.03 & IRCS  & SA97-249, LTT12162\\
1301 & Yvonne     & 2013.04.06 & 12:46--15:23 & 3.189 & 2.674 & 16.9 & 1.01--1.12 & IRCS  & SA102-1081\\
1764 & Cogshall   & 2013.12.06 & 14:02--16:03 & 3.462 & 2.494 & 3.6  & 1.05--1.12 & KOOLS & HD28099, HD89010\\
2250 & Stalingrad & 2013.12.05 & 10:34--11:26 & 2.814 & 2.169 & 17.3 & 1.16--1.19 & KOOLS & HD1368\\
2310 & Olshaniya  & 2013.12.06 & 09:25--10:44 & 2.934 & 2.459 & 18.5 & 1.23--1.24 & KOOLS & HD1368, SA93-101, SA98-978, SA102-1081\\
2376 & Martynov   & 2013.12.08 & 17:02--18:38 & 3.263 & 2.468 & 11.8 & 1.02--1.05 & KOOLS & HD25680, HD65523, HD70088\\
2519 & Annagerman & 2012.02.16 & 11:45--12:13 & 3.466 & 2.644 & 10.4 & 1.09--1.13 & IRCS  & SA103-272, HD127913\\ 
2525 & O'Steen    & 2012.02.16 & 12:47--13:28 & 3.725 & 3.122 & 13.2 & 1.16--1.23 & IRCS  & SA103-272, WASP-1\\
2667 & Oikawa     & 2013.12.06 & 12:03--12:49 & 2.735 & 1.955 & 15.0 & 1.10--1.11 & KOOLS & HD28099, HD89010\\
2670 & Chuvashia  & 2013.12.06 & 18:05--18:53 & 3.020 & 2.887 & 19.0 & 1.37       & KOOLS & HD1368\\
     &            & 2013.12.08 & 19:12--20:40 & 3.021 & 2.860 & 19.0 & 1.26--1.38 & KOOLS & HD1368, SA102-1081\\ 
     &            & 2014.01.28 & 17:42--18:01 & 3.053 & 2.253 & 12.6 & 1.31       & KOOLS & SA102-1081\\
3037 & Alku       & 2013.12.06 & 18:05--18:53 & 2.213 & 1.500 & 21.5 & 1.05--1.06 & KOOLS & HD28099, HD89010\\
3104 & Durer      & 2013.04.06 & 08:18--10:11 & 3.120 & 2.236 & 10.2 & 1.01--1.09 & IRCS  & SA102-1081\\
4896 & Tomoegozen & 2013.12.08 & 10:31--11:41 & 2.655 & 2.053 & 19.2 & 1.03--1.05 & KOOLS & HD25680, HD65523, HD70088\\ 
4925 & Zhoushan   & 2013.12.08 & 12:18--15:37 & 2.454 & 1.481 & 4.5  & 1.04--1.14 & KOOLS & HD25680, HD65523, HD70088\\ 
\end{longtable}
}
\end{landscape}
\footnotetext[3]{$\langle$http://ssd.jpl.nasa.gov/horizons.cgi\#top$\rangle$.}

\subsection{Near-infrared spectroscopic observations}
Spectroscopic data in near-infrared wavelength regions were obtained using the infrared camera and spectrograph (IRCS; \cite{Tokunaga1998}; \cite{Kobayashi2000}) attached to the f/12 Cassegrain focus of the 8.2 m Subaru Telescope.
The instrument detector has a 1024 $\times$ 1024 Aladdin-III InSb array, which provides a \timeform{54''} $\times$ \timeform{54''} field-of-view, with a pixel scale of \timeform{0''.052}.
The IRCS system is composed of a number of grisms. 
Spectroscopic data are obtained by dividing the spectral range into two wavelength regions: \textit{JH} and \textit{HK}.
The two wavelength ranges determined using the \textit{JH} grism with the order-sorting filter for \textit{JH} mode and the \textit{HK} grism with the order-sorting filter for \textit{HK} mode were 1.0--1.6 \micron\ and 1.45--2.5 \micron\, respectively.
There is an overlap wavelength of the two grism modes in 1.45--1.60 \micron.
The IRCS in low resolution has a dispersion of approximately 13 \AA\ per pixel in the \textit{JH} grism, and approximately 12 \AA\ per pixel in the \textit{HK} band grism, in the wavelength direction.
The orientation of the slit of most observations was along the parallactic angle to prevent the flux loss caused by atmospheric dispersion.
To increase the signal-to-noise ratio and avoid the flux loss due to the slit, a 188-element curvature sensor adaptive optics (AO188) system \citep{Minowa2010} was used with IRCS observations.
Since the target asteroids were too faint, the nearby bright stars were used as the natural guide star to operate the AO188 system.
Typical seeing size with the AO188 system was improved to $\sim$ \timeform{0''.2}--\timeform{0''.6} in \textit{K} band through the observations.
\timeform{0''.3}, \timeform{0''.6}, and \timeform{0''.9}-wide slits were used according to the changes in the seeing size.
The length of each width component of the long slit for IRCS was $\sim$ \timeform{15''} in the cross-wavelength direction.
The spectral resolution in low dispersion in \textit{JH} and \textit{HK} grisms was $\sim$100 \AA\, in case of using the \timeform{0''.9} slit.

\subsection{Visible spectroscopic observations}
Optical spectroscopic data were recorded using the Kyoto Okayama Optical Low-dispersion Spectrograph (KOOLS; \cite{Ohtani1998}; \cite{Ishigaki2004}) installed at the f/18 Cassegrain focus of the 1.88 m telescope at the Okayama Astrophysical Observatory.
The 2048 $\times$ 4096 SITe ST-002A CCD installed on KOOLS with 15 \micron\ square pixels, produces a \timeform{5'} $\times$ \timeform{4'.4} field-of-view with a pixel scale of \timeform{0''.3}.
A grism with the 6563 \AA\ $\mathrm{mm^{-1}}$ blaze and a Y49 order-sorting filter were used, and the covered spectral range is 0.47--0.88 \micron.
The KOOLS has a dispersion of approximately 3.8 \AA\ per pixel in the wavelength direction.
The length of the long slit for KOOLS was \timeform{4'.4} in the cross-wavelength direction.
Since the 1.88 m telescope did not have an atmospheric dispersion corrector that can prevent flux loss errors, it was necessary to select a wide slit for KOOLS observations; thus, a large width of \timeform{6''}, which yields a spectral resolution of 32 \AA\ was utilized.

\subsection{Methods of data acquisition for asteroids}
To acquire an asteroid image as a point source, non-sidereal trackings were employed for the Subaru and 1.88 m telescopes.
As a typical method to obtain the reflectance spectrum of the asteroid, the asteroid flux spectrum is divided by the solar flux spectrum.
Therefore, solar analogue stars classified as G or F8 stars in the SIMBAD Astronomica database\footnotemark[2] or \citet{Drilling1979} were used as standard stars.
Observations of the solar analogue stars were coordinated so that the airmass difference between the asteroid and the solar analogue star was less than 0.1 in each case.
Spectroscopic observations were performed with an airmass lower than 1.4 to prevent the airmass mismatch between the asteroid and solar analogue star.
The asteroids and solar analogue stars were situated in the center of the slit during spectroscopic observations to avoid flux envelope loss due to slits.
To reduce noise in signals due to atmospheric emission lines and thermal emissions, spectroscopic data were alternated between two different slit positions, ``A'' and ``B'' positions, on the detector.
An observation set for the asteroids was composed of four positions, ``A--B--B--A''.
The ``A--B--B--A'' sequence for IRCS and KOOLS observations was conducted using the two-position nodding patterns with an interval of \timeform{7''} and \timeform{20''} along the cross-wavelength direction of the slit, respectively.
The spectrum on the obtained image was recurrently located on the same column to ensure the reproducibility signal.
For IRCS and KOOLS observations, flat-fielding images were obtained with a halogen lamp.
Wavelength calibration frames for IRCS and KOOLS were acquired regularly at night with light from argon and iron--neon--argon hollow cathode lamps, respectively.
\footnotetext[2]{$\langle$http://simbad.u-strasbg.fr/simbad/$\rangle$.}

\subsection{Data reduction procedures}
Reduction procedures for all spectroscopic data were conducted using the image reduction and analysis facilities (IRAF) software.
For sensitivity correction of each pixel, each object frame for asteroids and solar analogue stars was divided by the flat-fielding flame.
The spectra were reduced by subtracting ``A'' and following ``B'' images to eliminate the sky background including the emission lines and thermal emissions.
The two-dimensional spectrum on the image was collapsed to one-dimensional spectrum using the \textit{apall} task with the IRAF software.
Next, wavelength reference positions were determined by adapting the wavelength calibration frame to the object frame.
The reflectance of the asteroid was calculated by dividing the spectrum of the asteroid into that of the solar analogue star.
Individual visible and near-infrared spectra were smoothed by taking the median of the final reflectance spectrum of each frame, which was normalized to 0.55 and 1.60 \micron\, respectively.

\section{Obtained spectra and taxonomic classification of asteroids}
\begin{figure*}
  \begin{center}
    \FigureFile(82mm,82mm){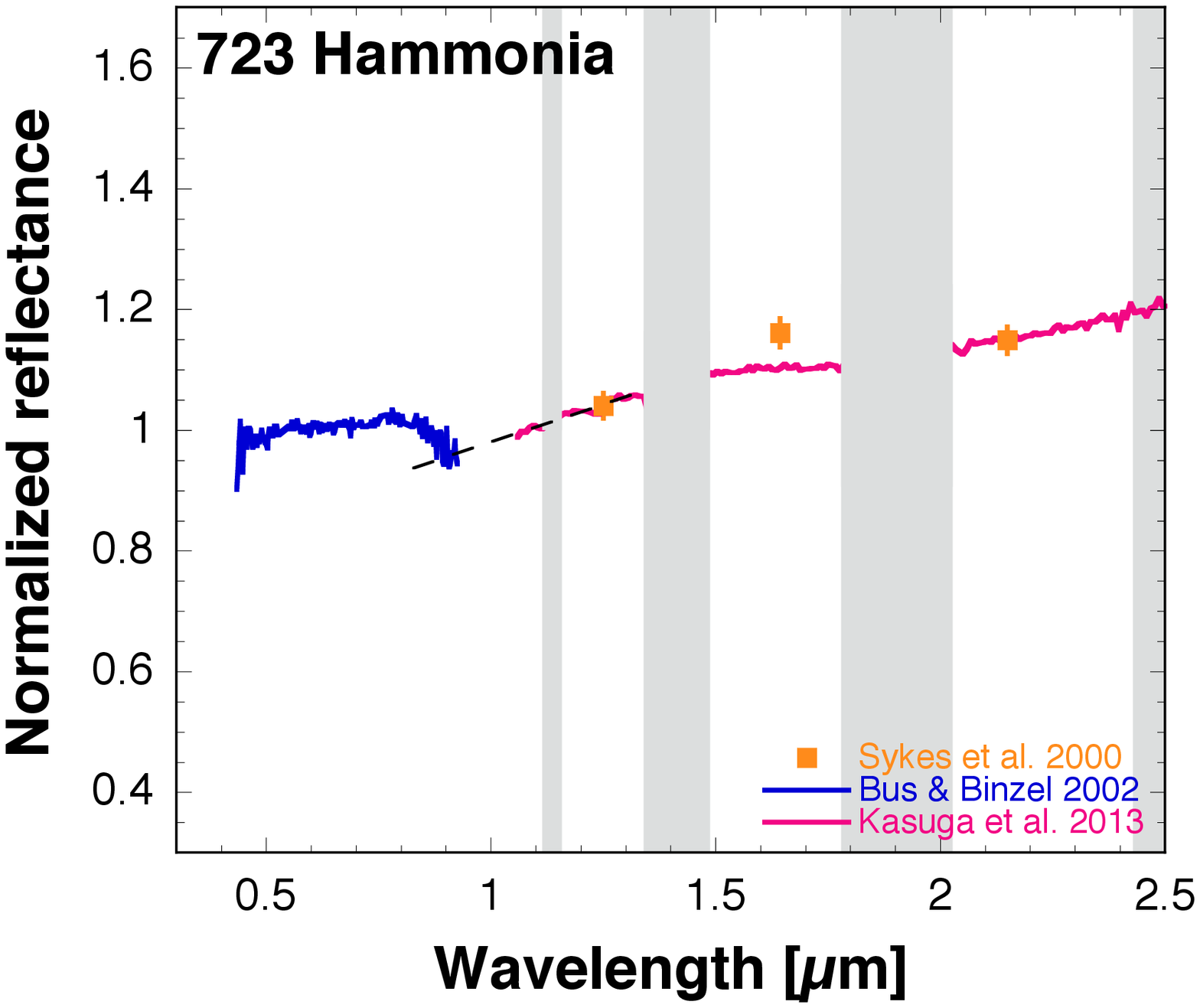}
    \FigureFile(82mm,82mm){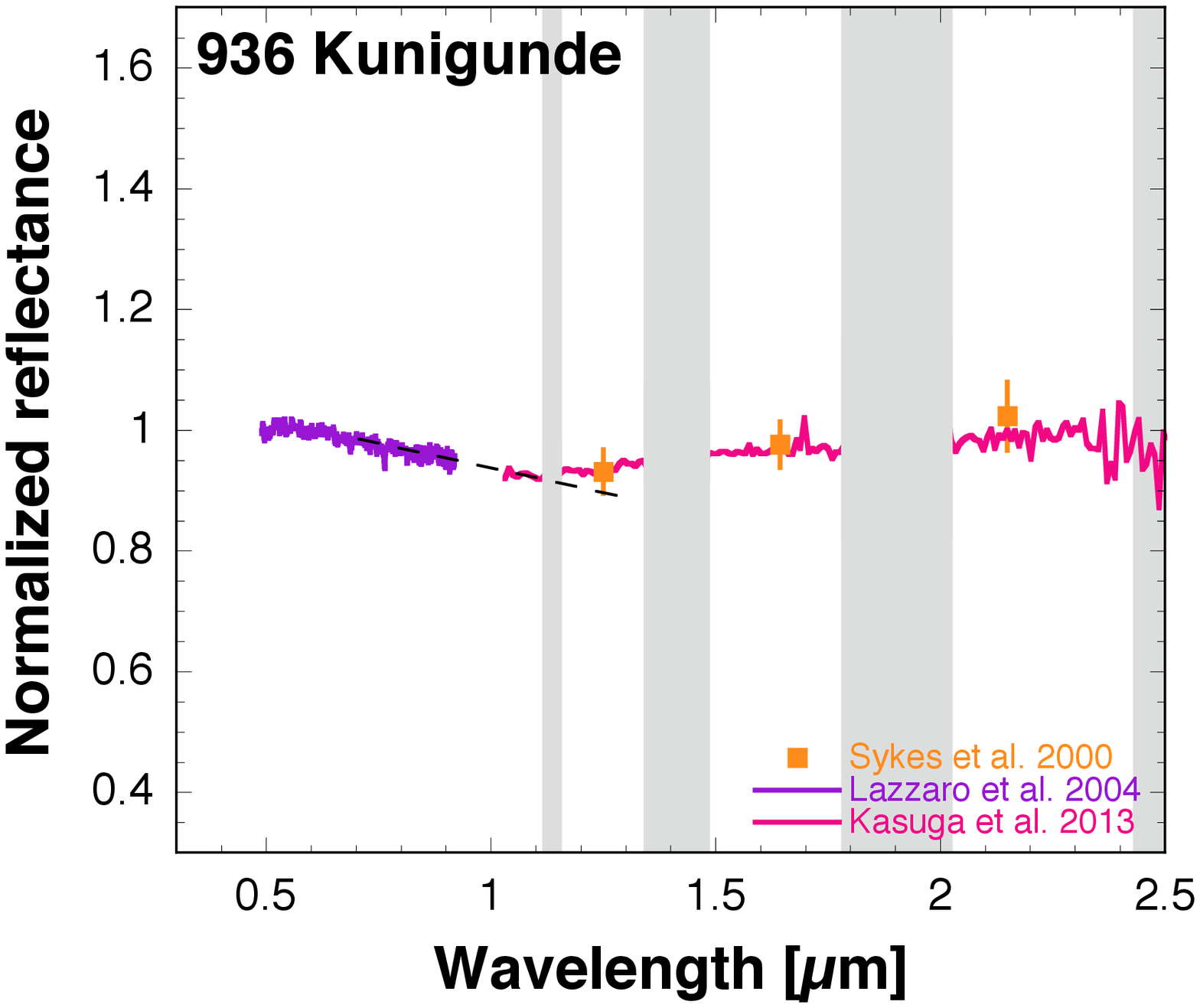}
    \FigureFile(82mm,82mm){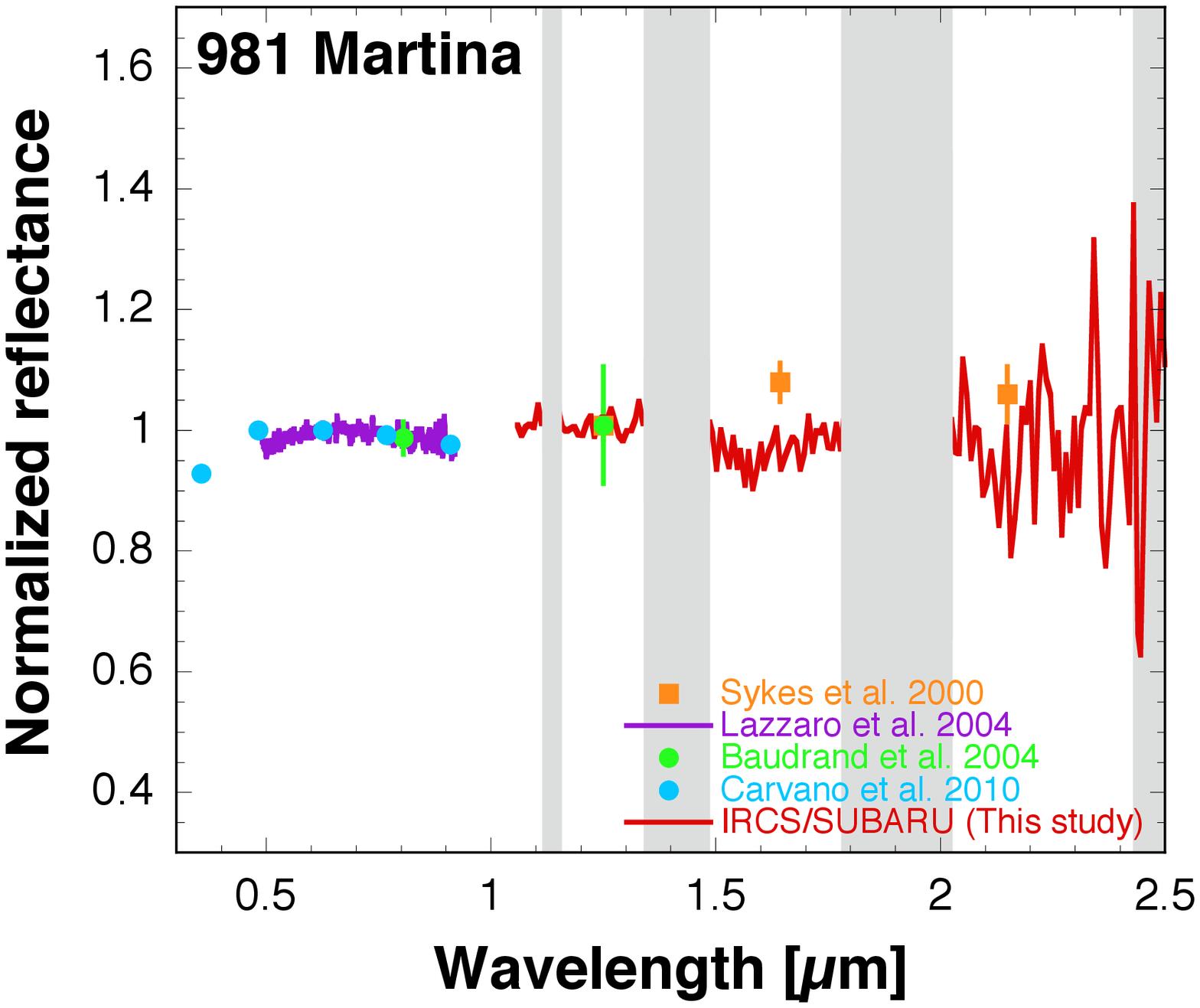}
    \FigureFile(82mm,82mm){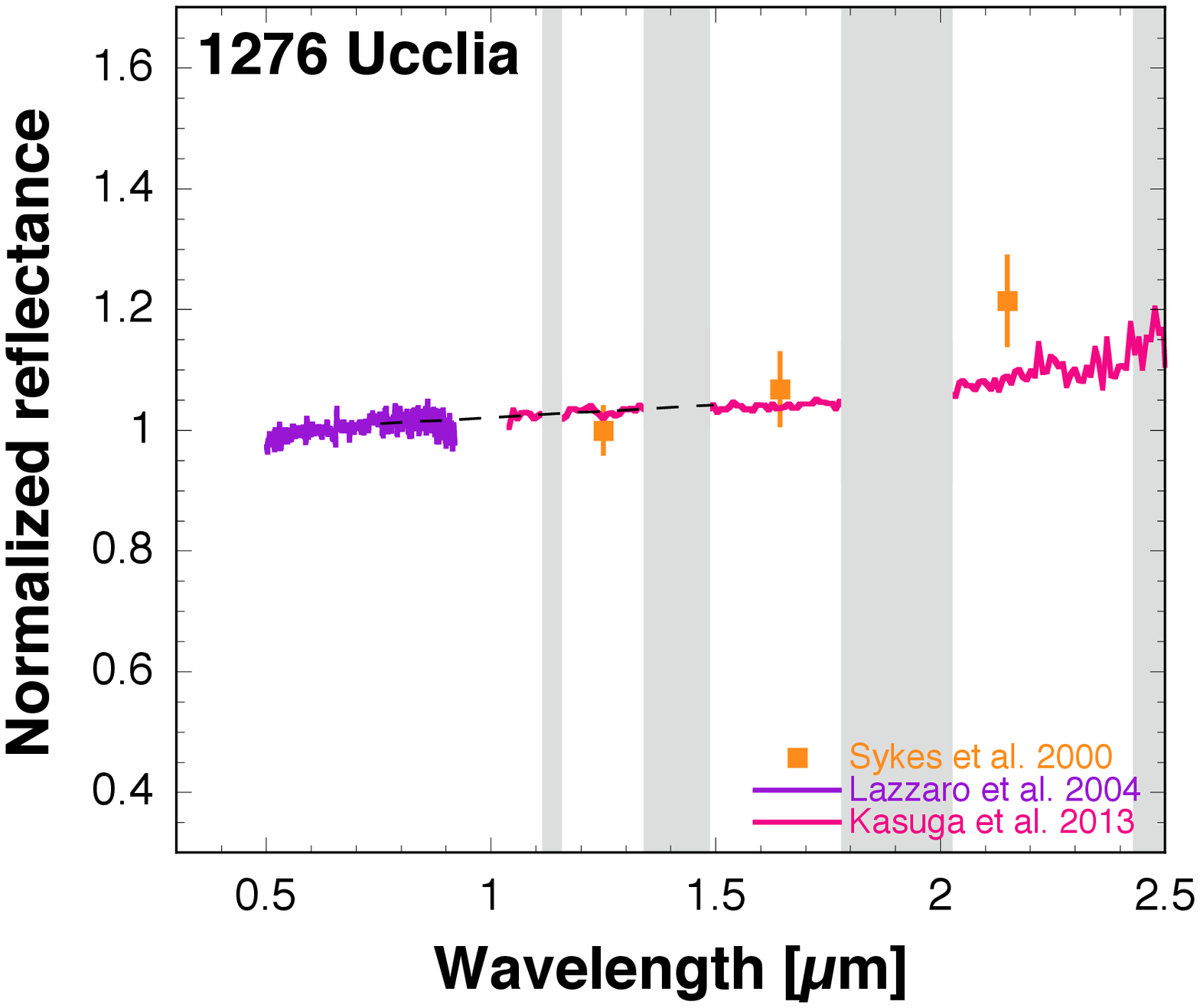}
    \FigureFile(82mm,82mm){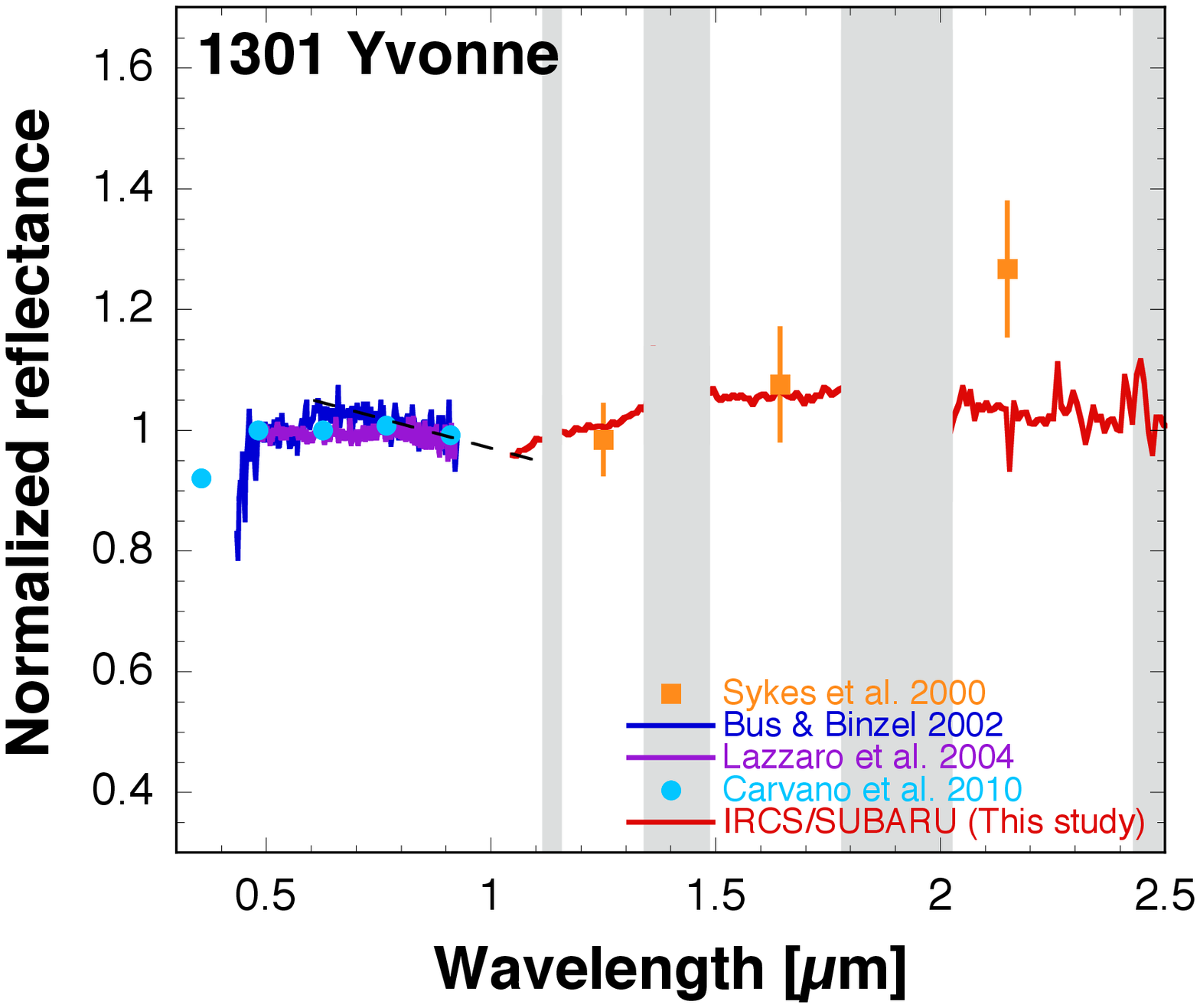}
    \FigureFile(82mm,82mm){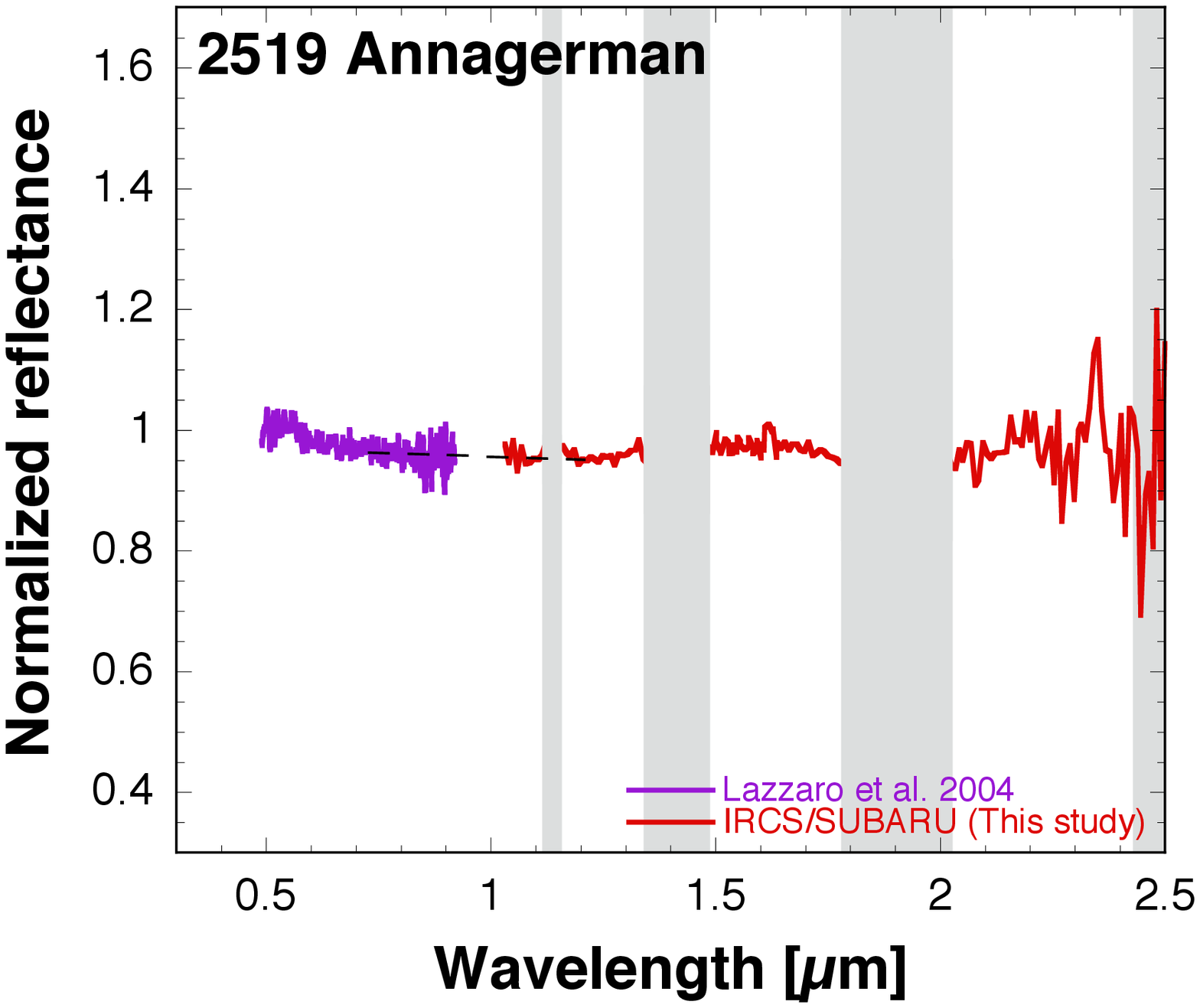}
  \end{center}
  \caption{Reflectance spectra of Bright C-complex and T-type asteroids in the Bus taxonomy.
The gray regions around 1.15, 1.40, 1.90, and 2.50 \micron\ indicate the wavelengths affected by the strong absorption due to the terrestrial atmosphere.
}
\label{fig:ast}
\end{figure*}

The spectra of the asteroids taken in this study are shown in figure \ref{fig:ast}--\ref{fig:ast3}.
The figures also show the spectra obtained in \authorcite{Kasuga2013} (\yearcite{Kasuga2013}, \yearcite{Kasuga2015}).
Spectroscopic and spectrophotometric data in visible and near-infrared wavelength regions reported in past studies (\cite{Zellner1985}; \cite{Sykes2000}\footnotemark[4]; \cite{Baudrand2001}; \cite{Bus2002a}; \cite{Baudrand2004}; \cite{Lazzaro2004}; \cite{Carvano2010}; \cite{Clark2010}; \cite{Fornasier2014}; \cite{Popescu2016}) are also displayed in figure \ref{fig:ast}--\ref{fig:ast3}.
Spectral data with wavelength gaps were interpolated to match both levels.
The dotted line in the figure is the line used for interpolation between the individual wavelength gap data.
\footnotetext[4]{The latest version is available from Sykes, M. V., Cutri, R., M., Skrutskie, M. F., Fowler, J. W., Tholen, D. J., Painter, P. E., Nelson, B., \& Kirkpatrick, D. J. 2010, NASA Planetary Data System, EAR-A-I0054/I0055-5-2MASS-V2.0.}

\addtocounter{figure}{-1}

\begin{figure*}
  \begin{center}
    \FigureFile(82mm,82mm){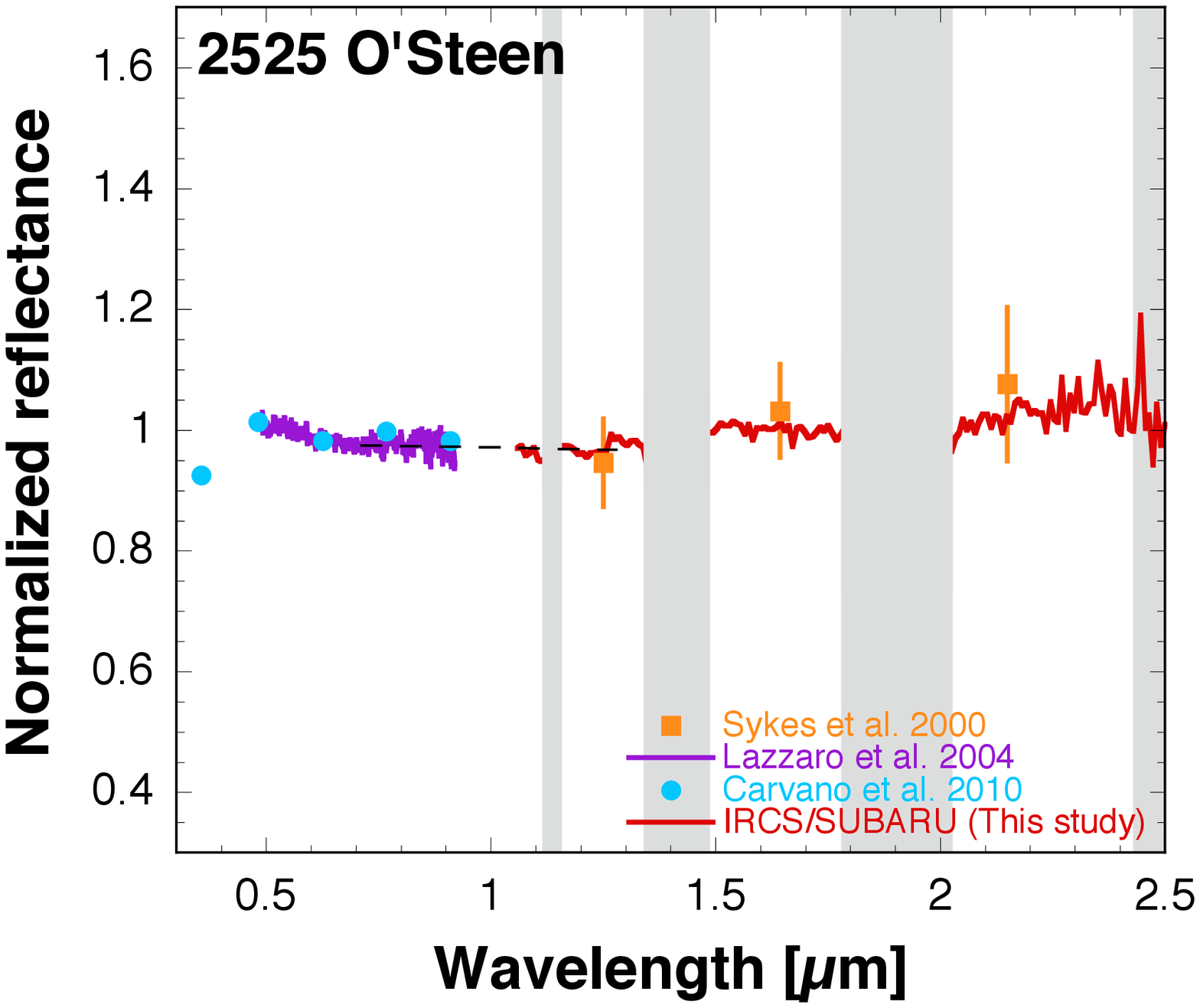}
    \FigureFile(82mm,82mm){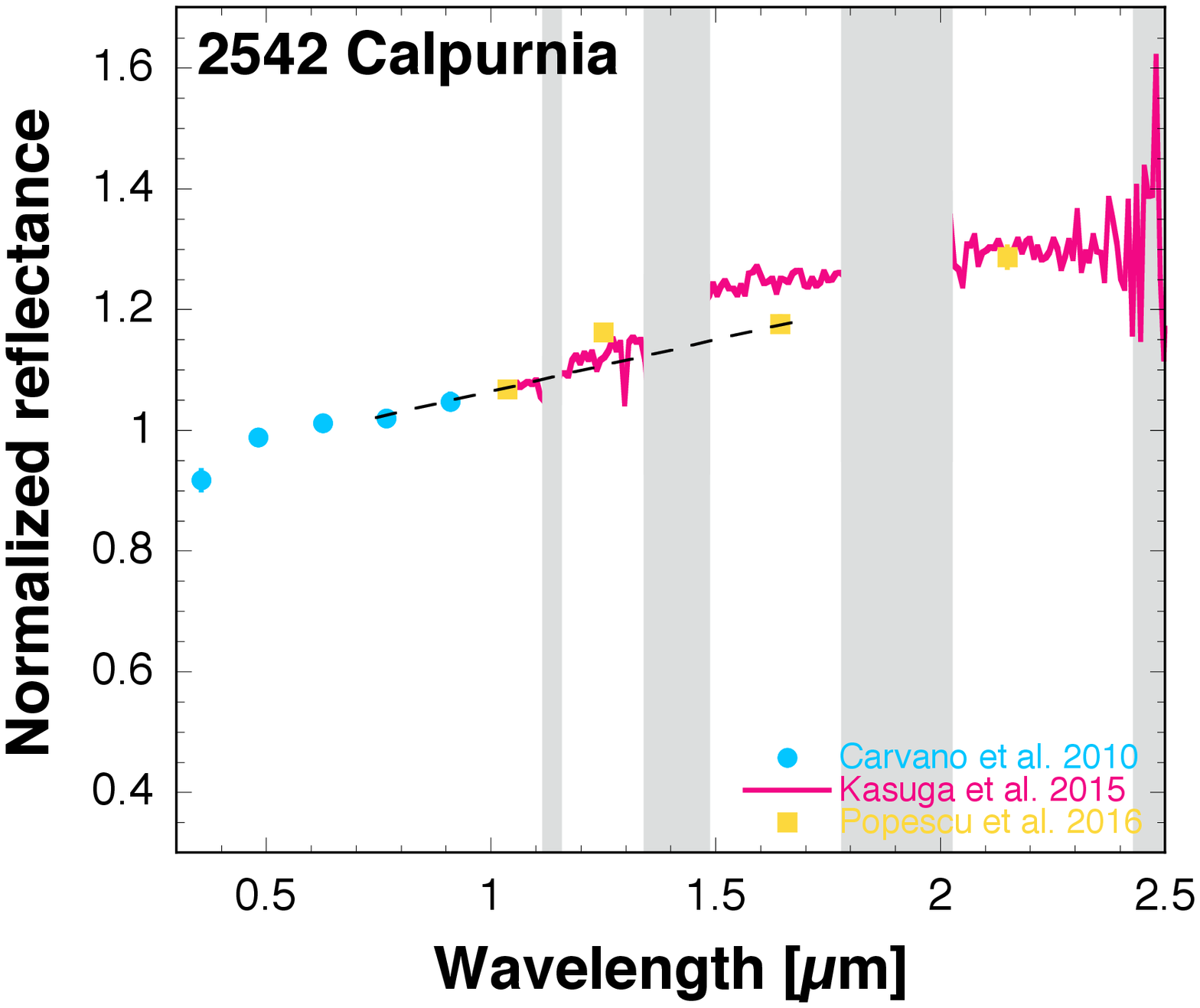}
    \FigureFile(82mm,82mm){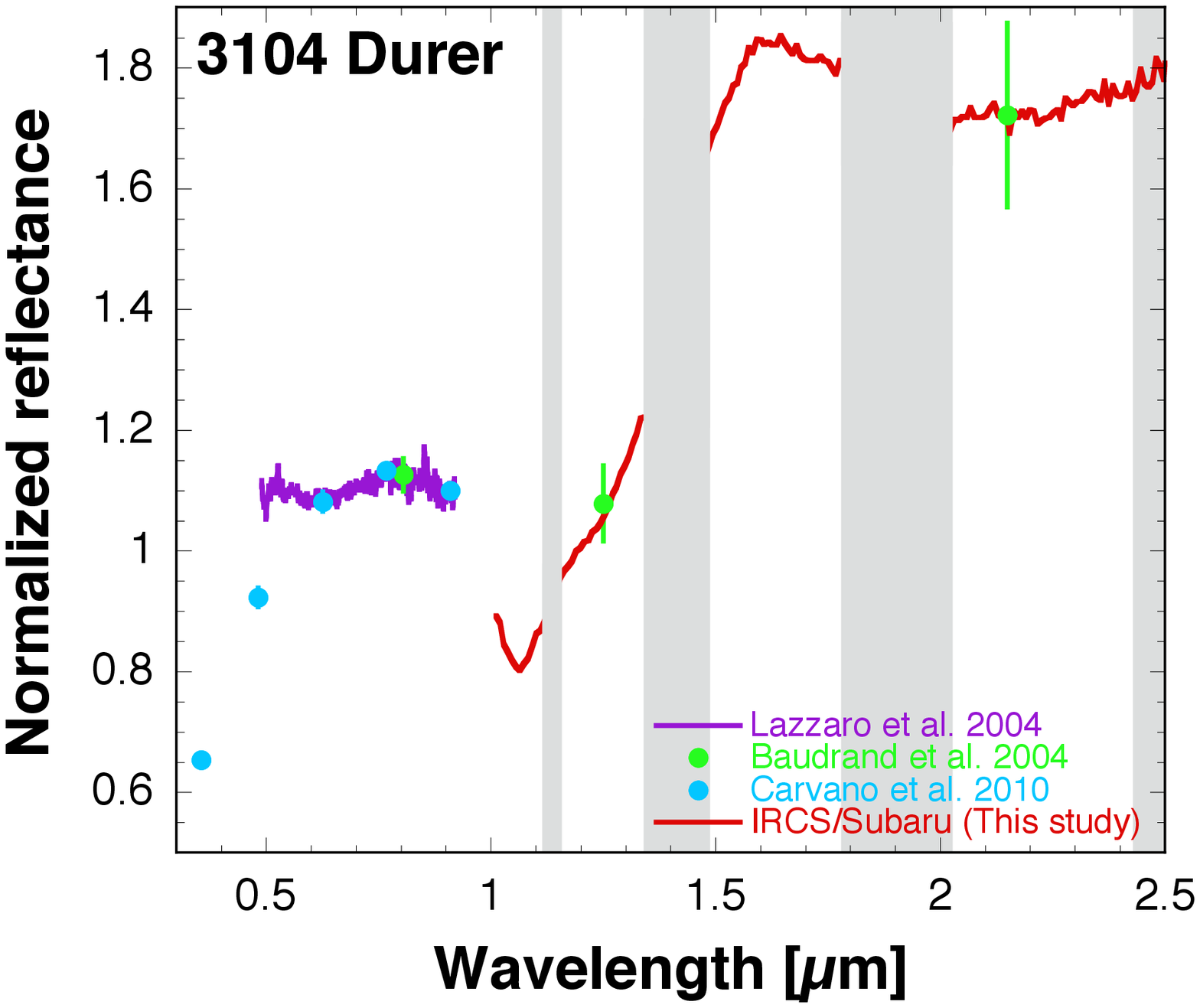}
    \FigureFile(82mm,82mm){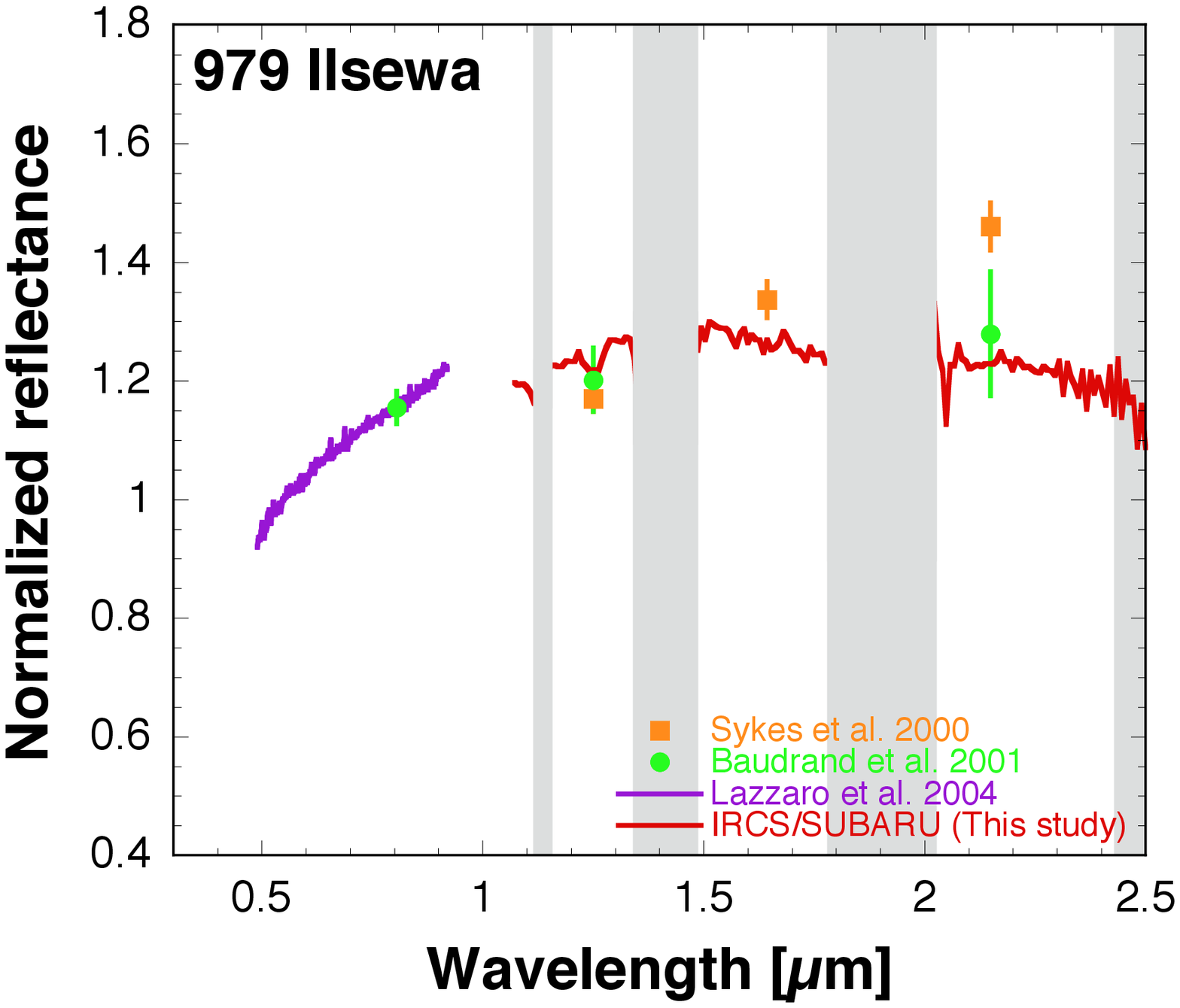}
  \end{center}
  \caption{\textcolor{red}{Continued}}
\label{fig:ast1d}
\end{figure*}

\begin{figure*}
  \begin{center}
    \FigureFile(82mm,82mm){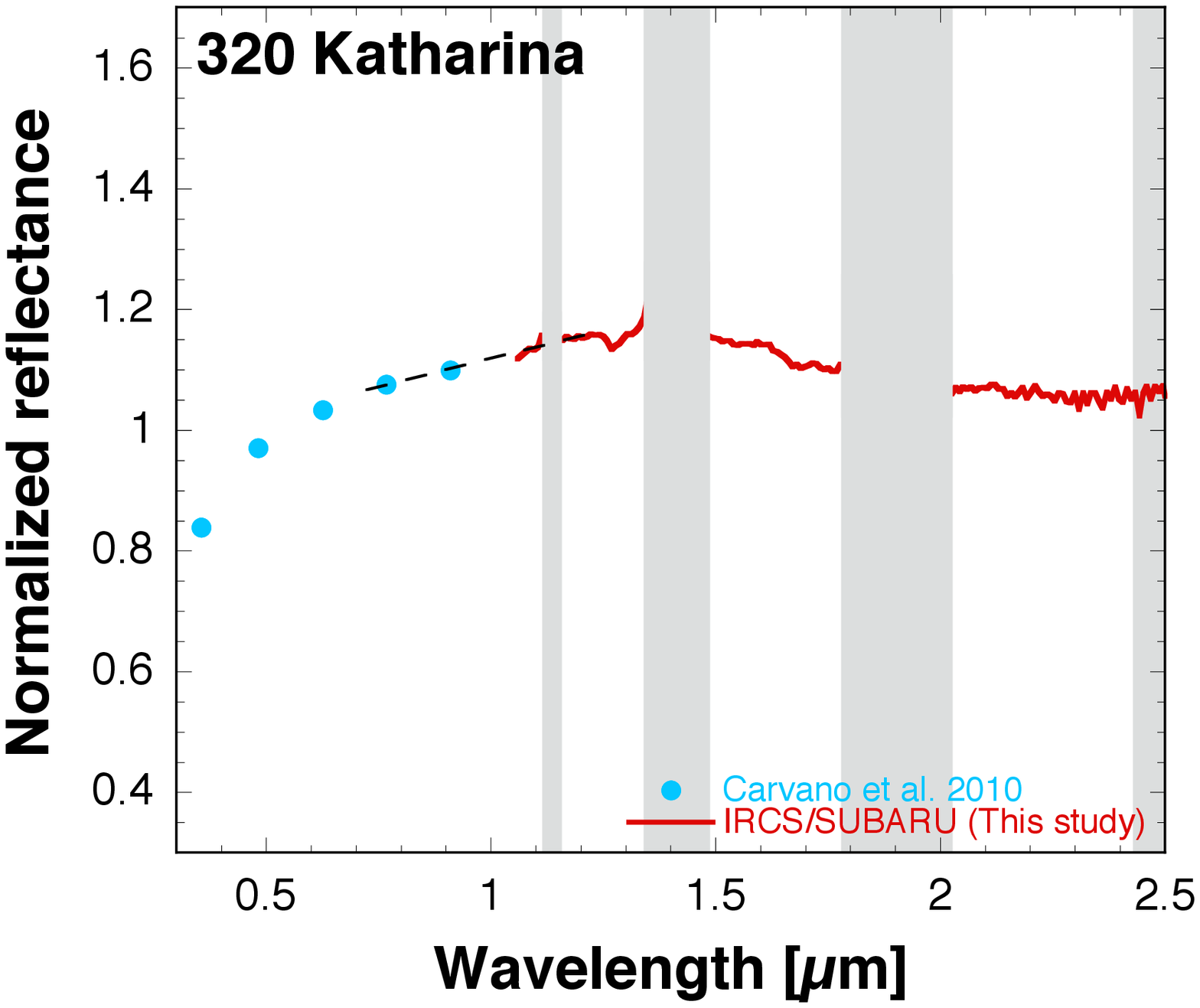}
    \FigureFile(82mm,82mm){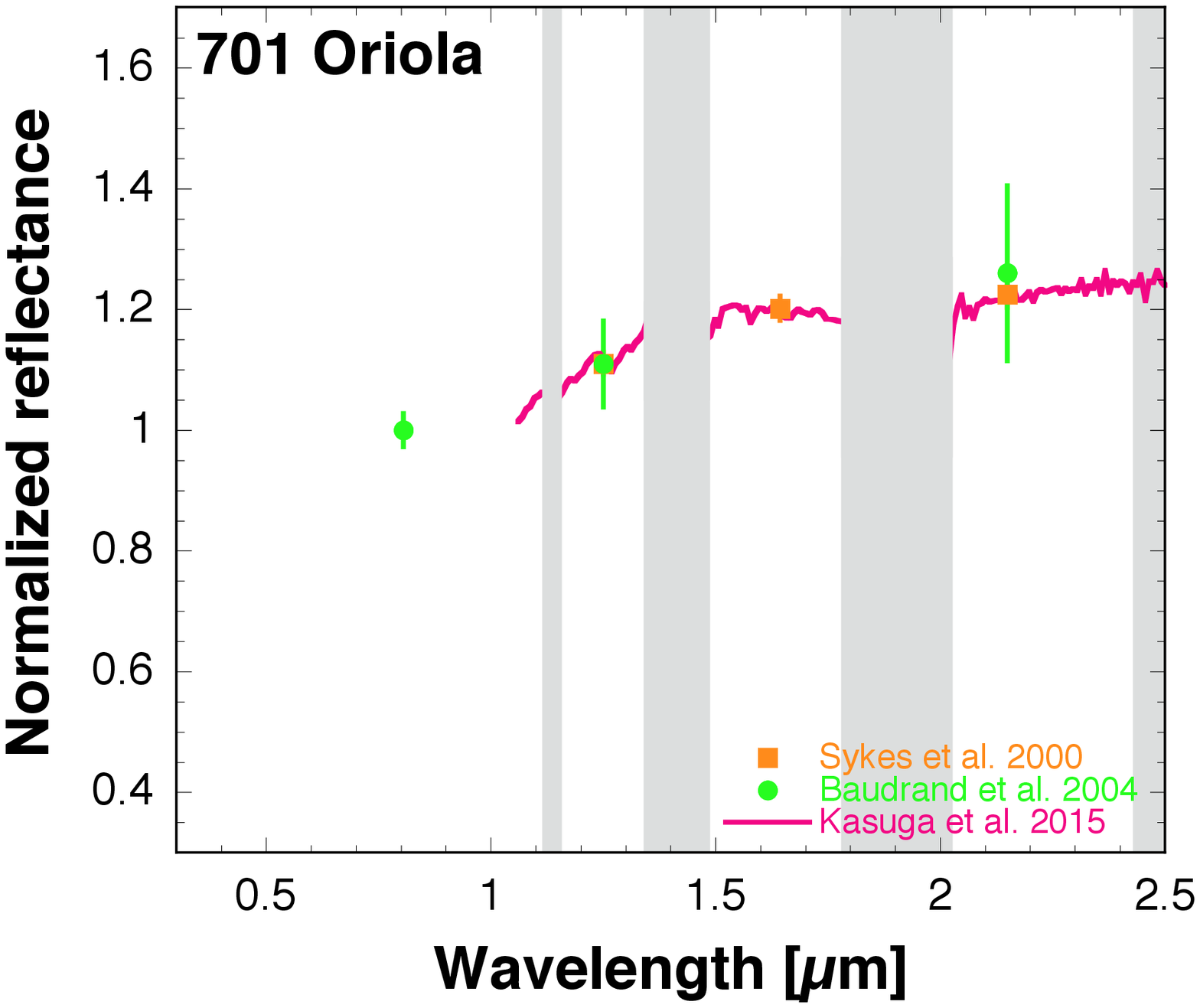}
    \FigureFile(82mm,82mm){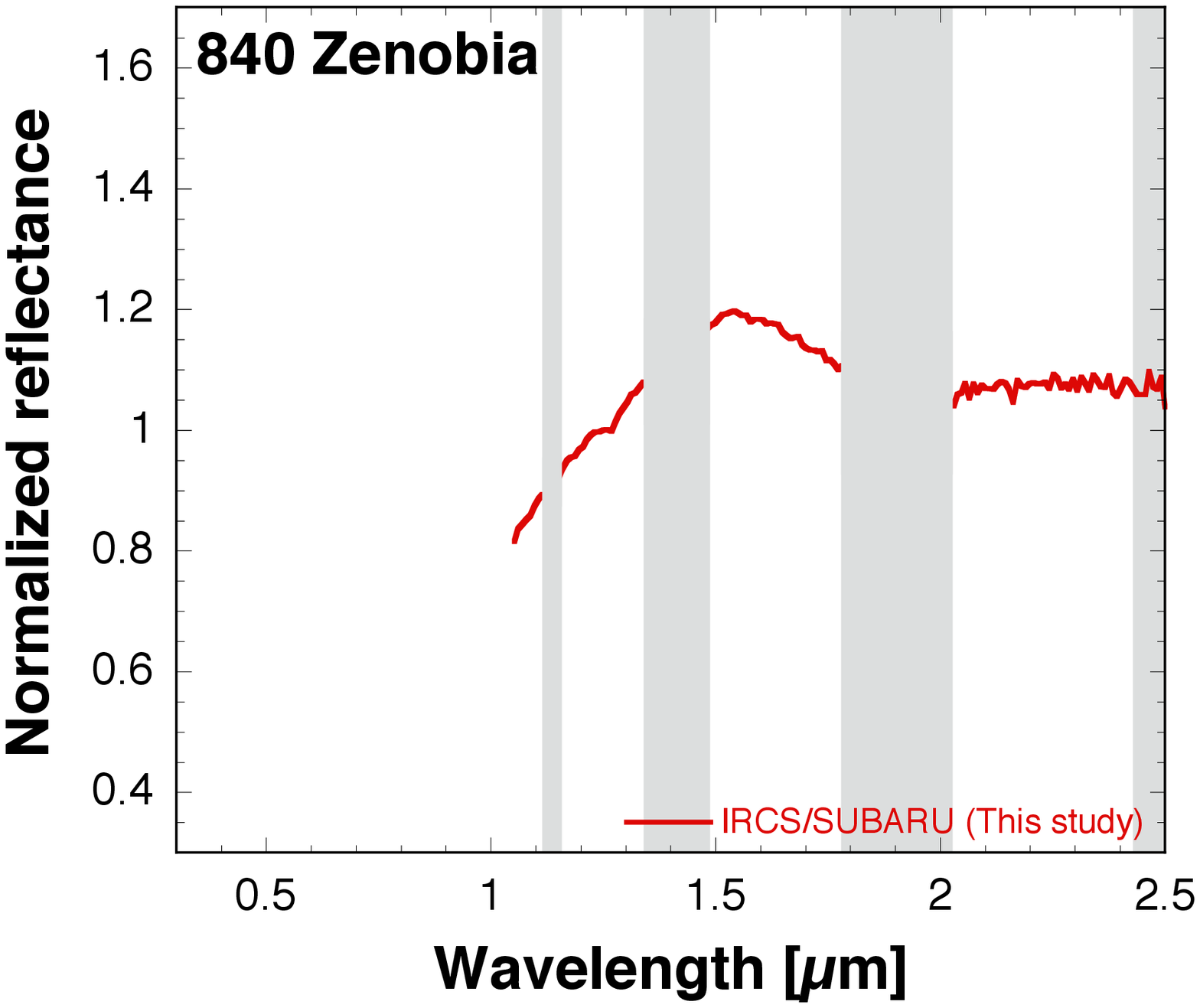}
    \FigureFile(82mm,82mm){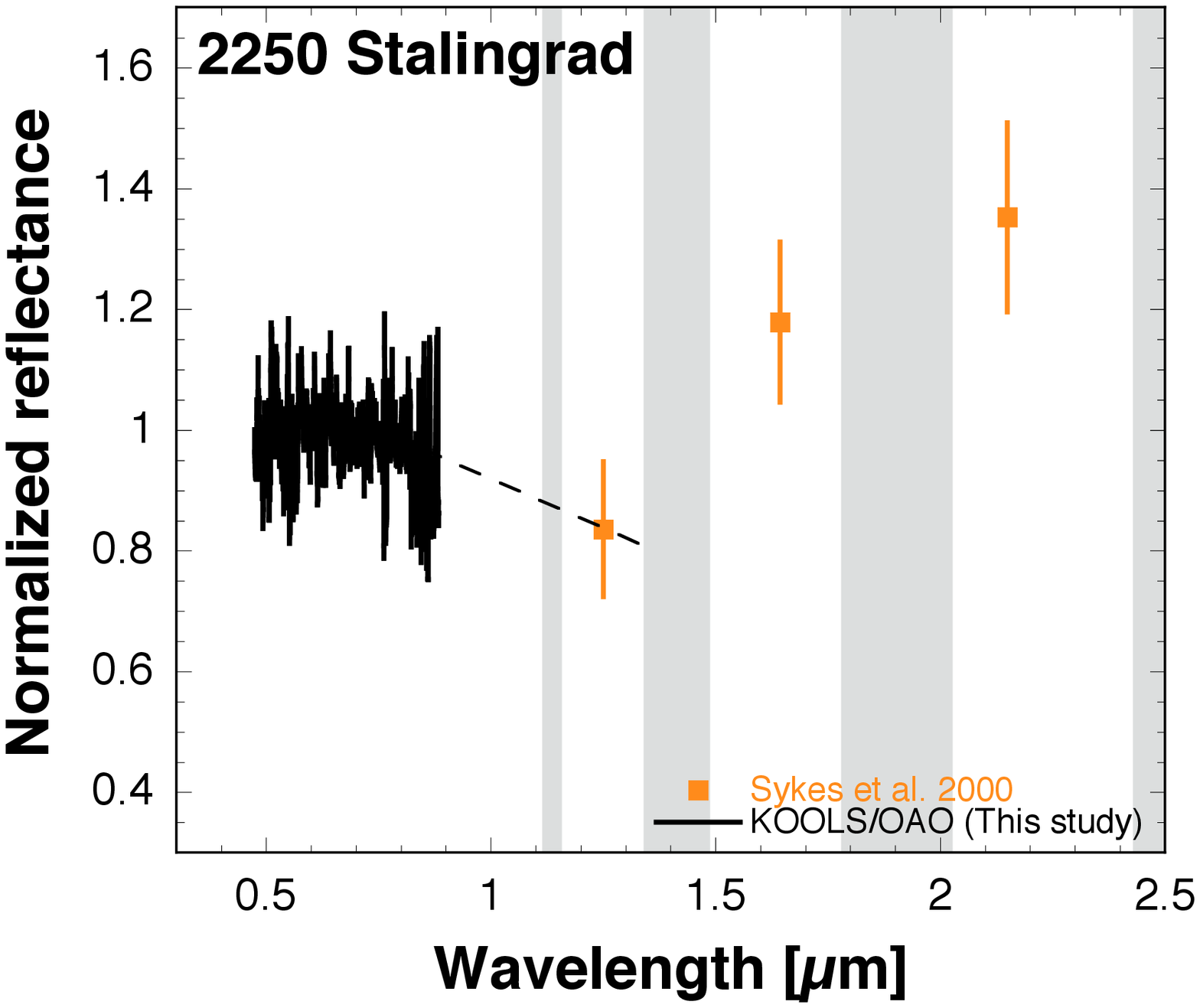}
    \FigureFile(82mm,82mm){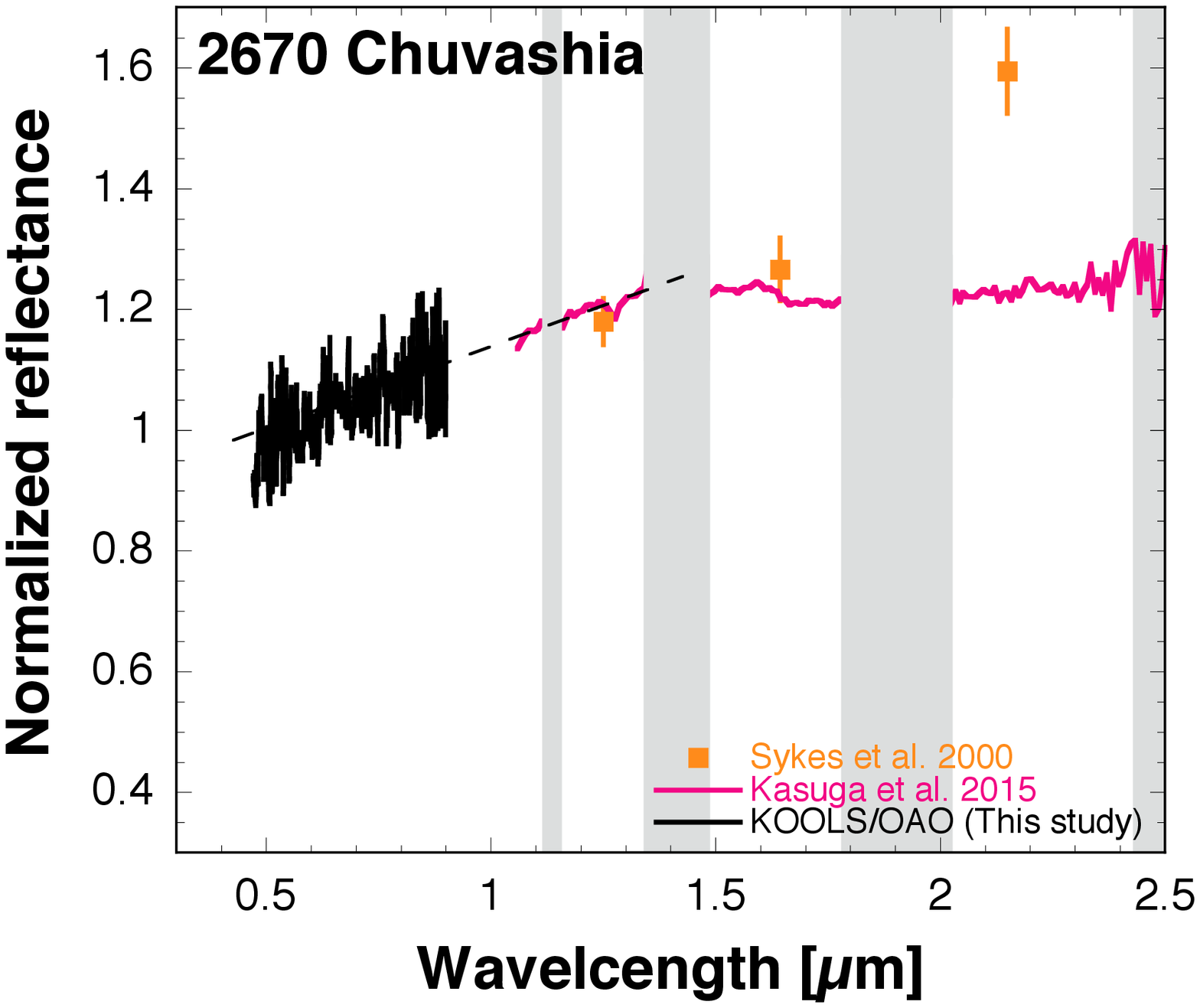}
  \end{center}
  \caption{Same as figure \ref{fig:ast} but for unknown spectral type asteroids with high albedo in the Bus taxonomy.
The reflectance is normalized at 0.55 \micron\ except in the case of asteroid 701 Oriola and 840 Zenobia.
The reflectance of asteroid 701 Oriola and 840 Zenobia are normalized at 0.8 and 1.25 \micron\, respectively.
}
\label{fig:ast2}
\end{figure*}
\addtocounter{figure}{-1}

\begin{figure*}
  \begin{center}
    \FigureFile(82mm,82mm){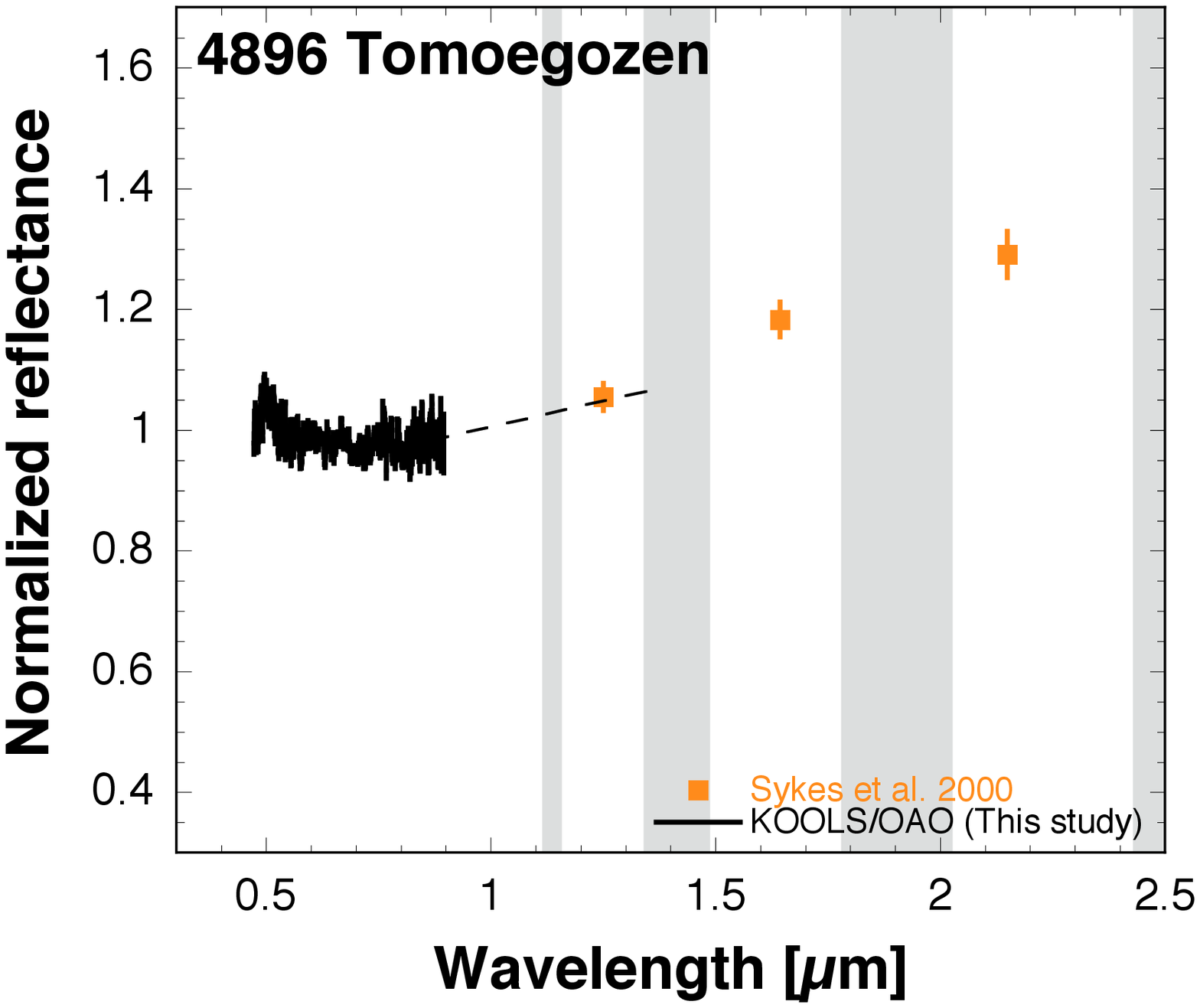}
    \FigureFile(82mm,82mm){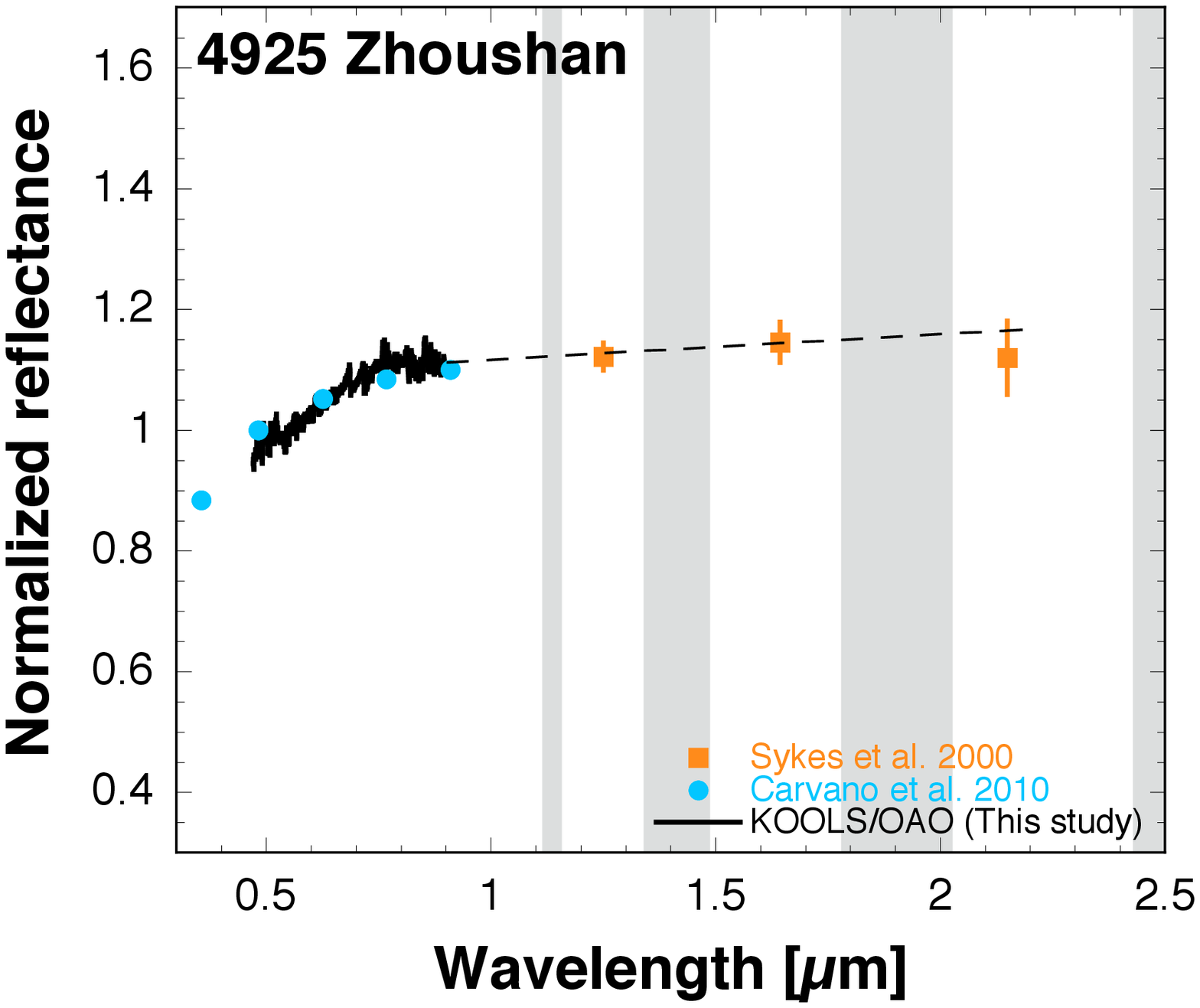}
  \end{center}
  \caption{Continued}
\label{fig:ast2d}
\end{figure*}

\begin{figure*}
  \begin{center}
    \FigureFile(82mm,82mm){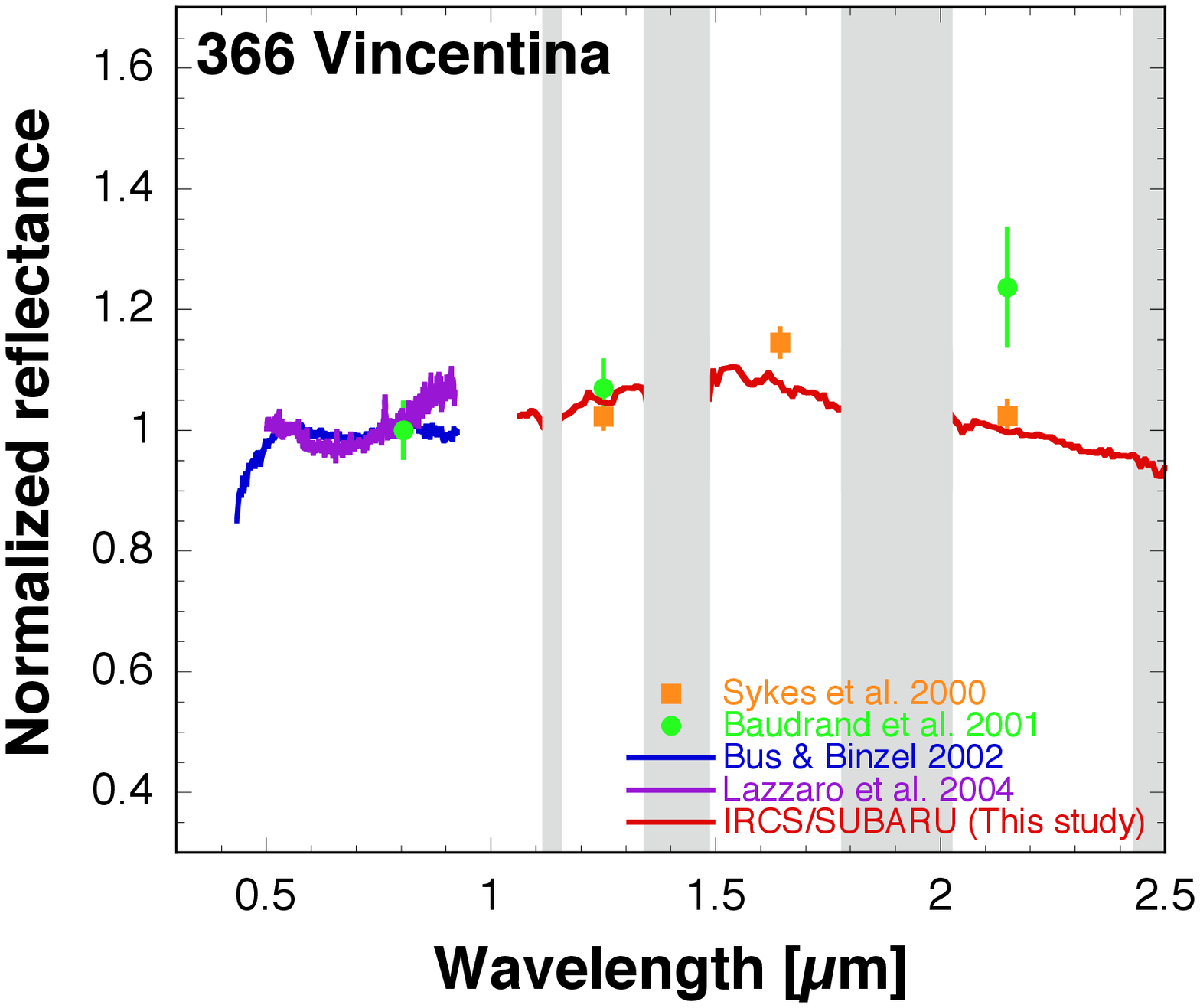}
    \FigureFile(82mm,82mm){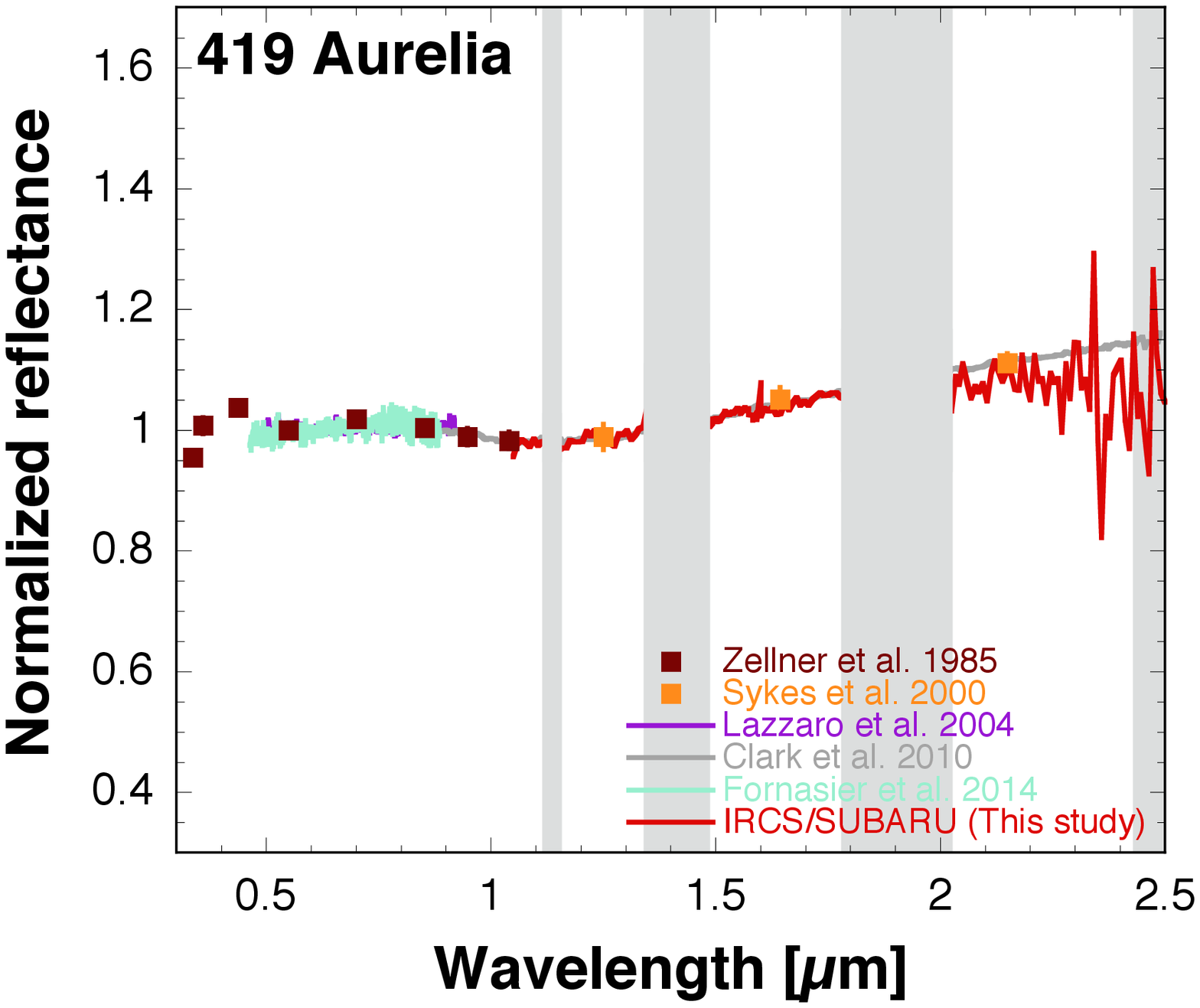}
    \FigureFile(82mm,82mm){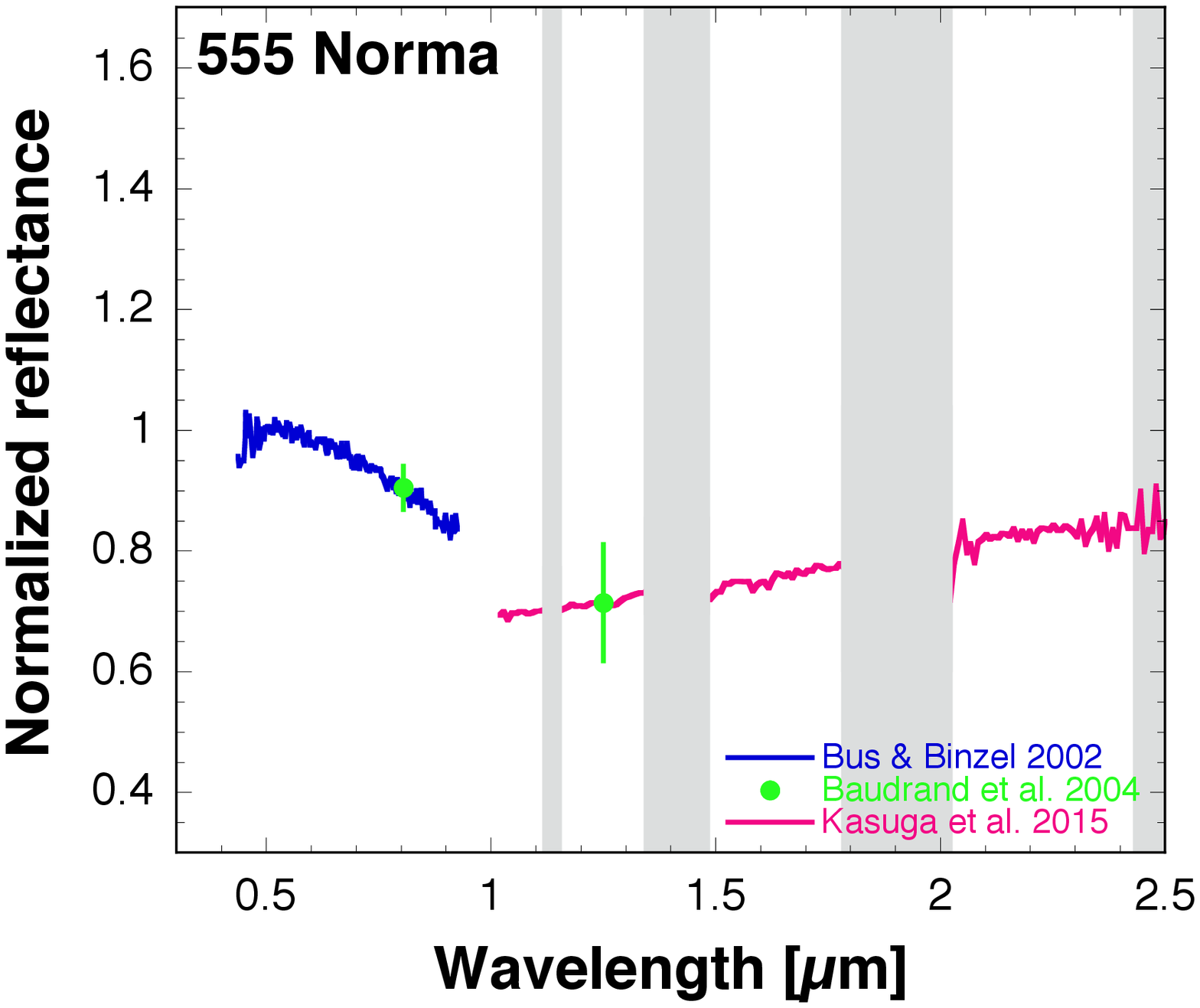}
    \FigureFile(82mm,82mm){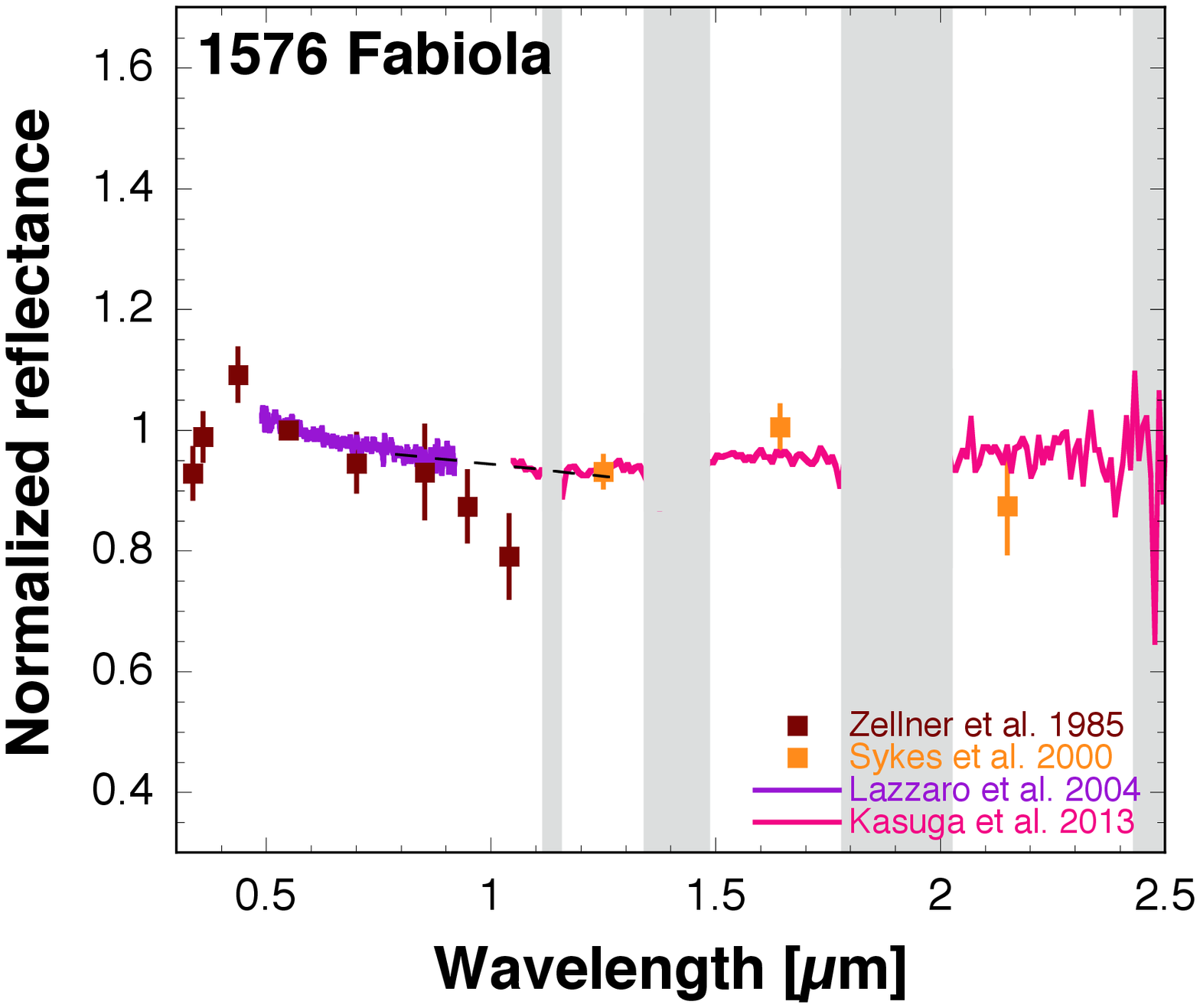}
    \FigureFile(82mm,82mm){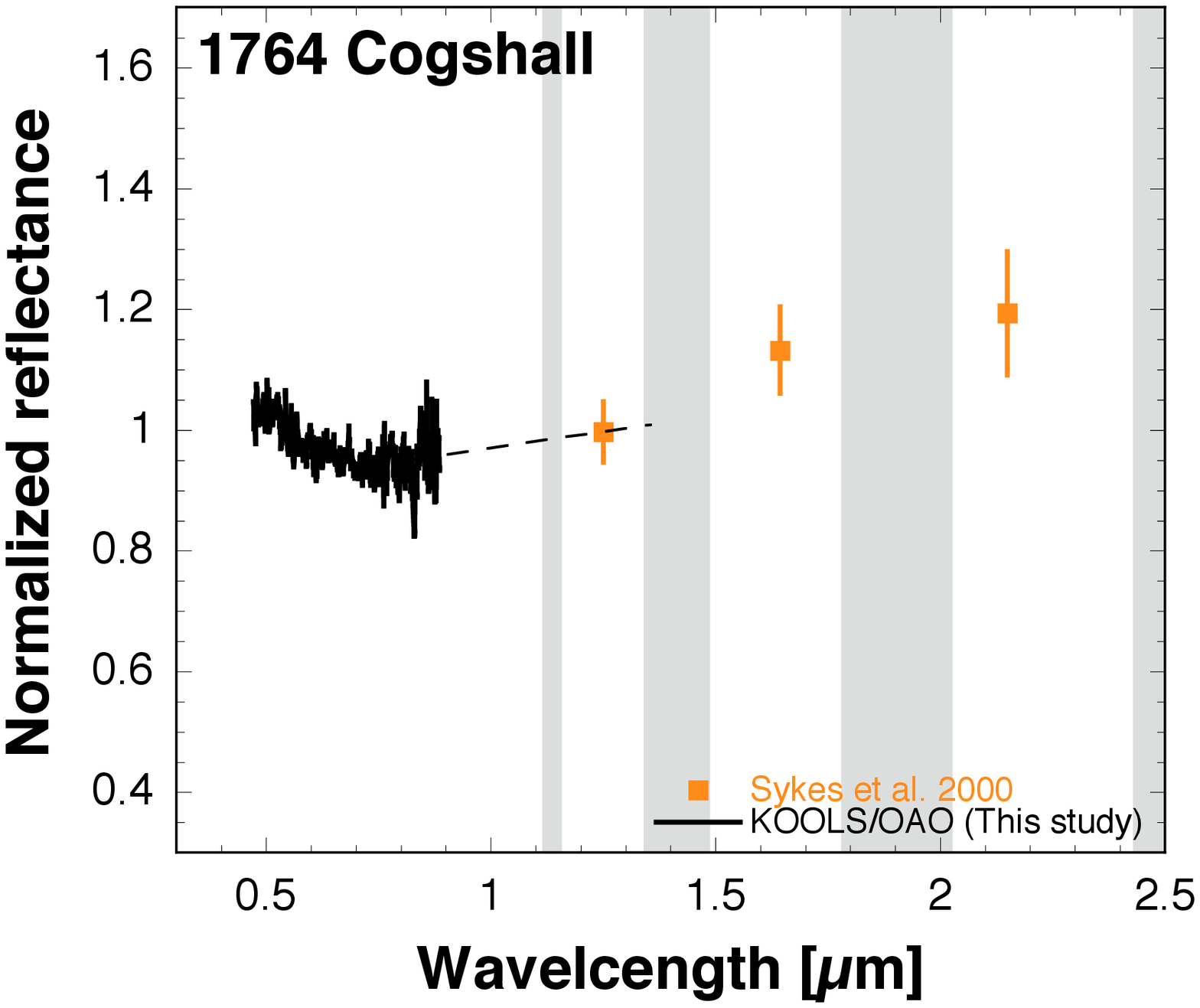}
    \FigureFile(82mm,82mm){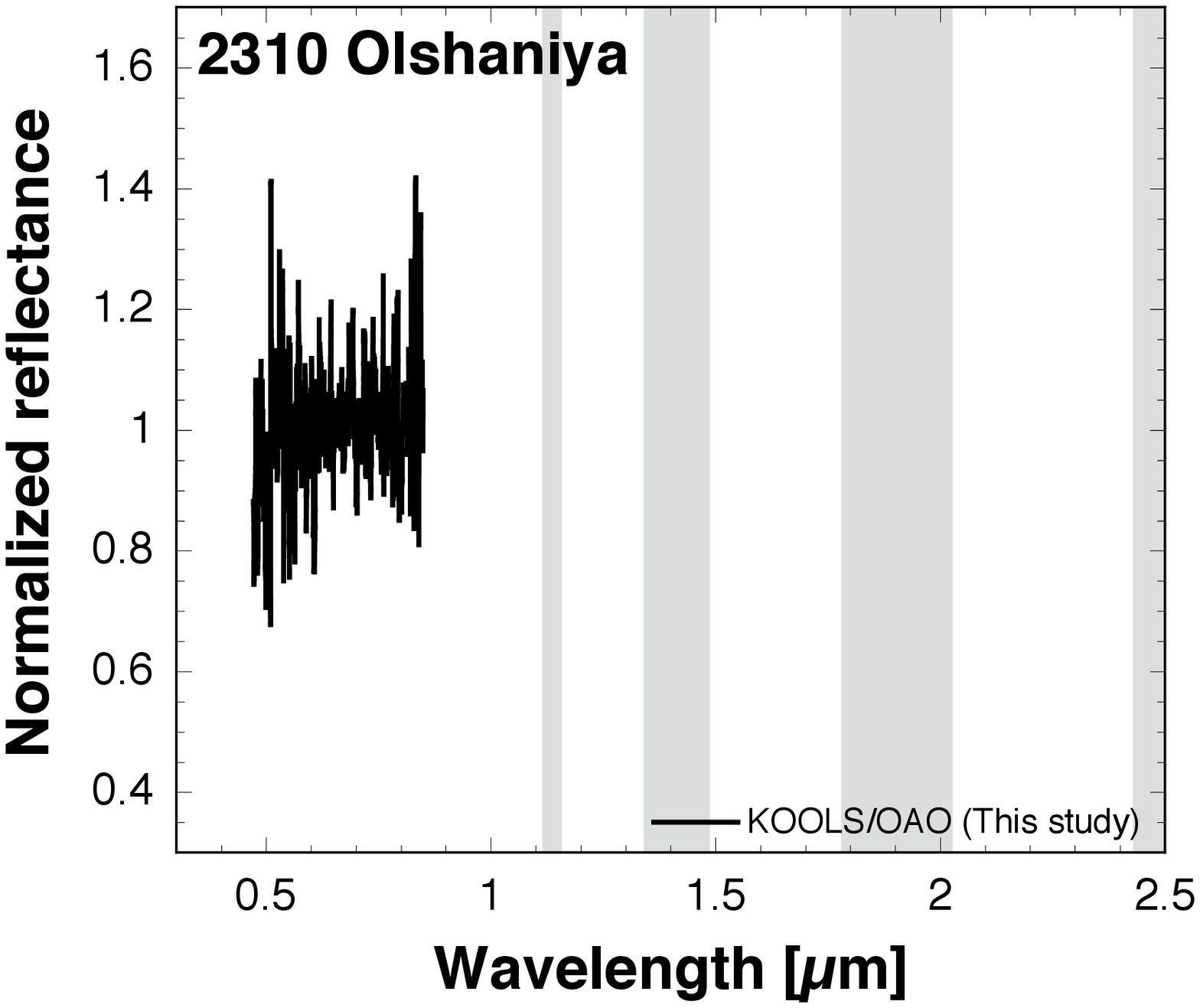}
  \end{center}
  \caption{Same as figure \ref{fig:ast} but for asteroids with low albedo.
}
\label{fig:ast3}
\end{figure*}
\addtocounter{figure}{-1}

\begin{figure*}
  \begin{center}
    \FigureFile(82mm,82mm){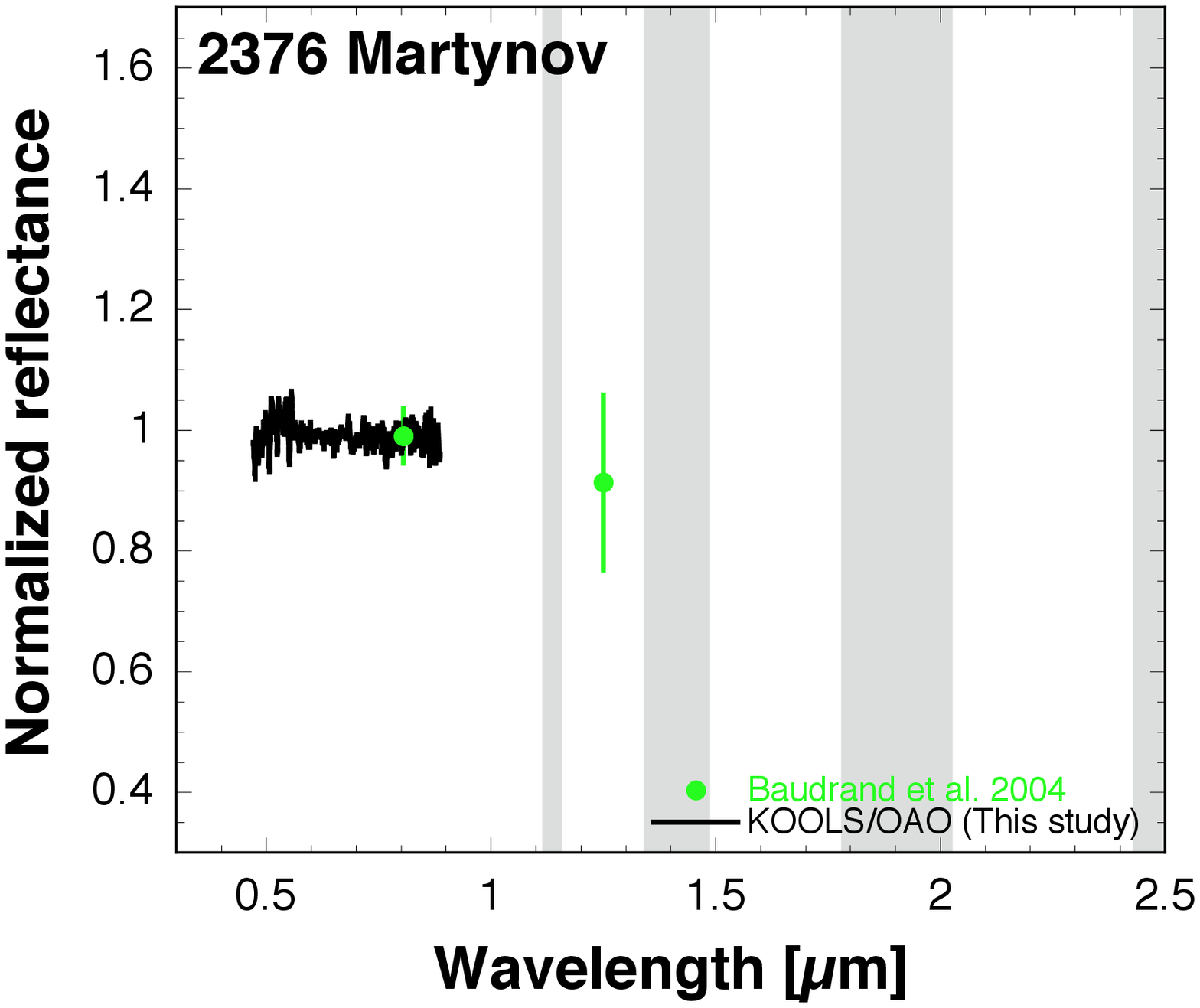}
    \FigureFile(82mm,82mm){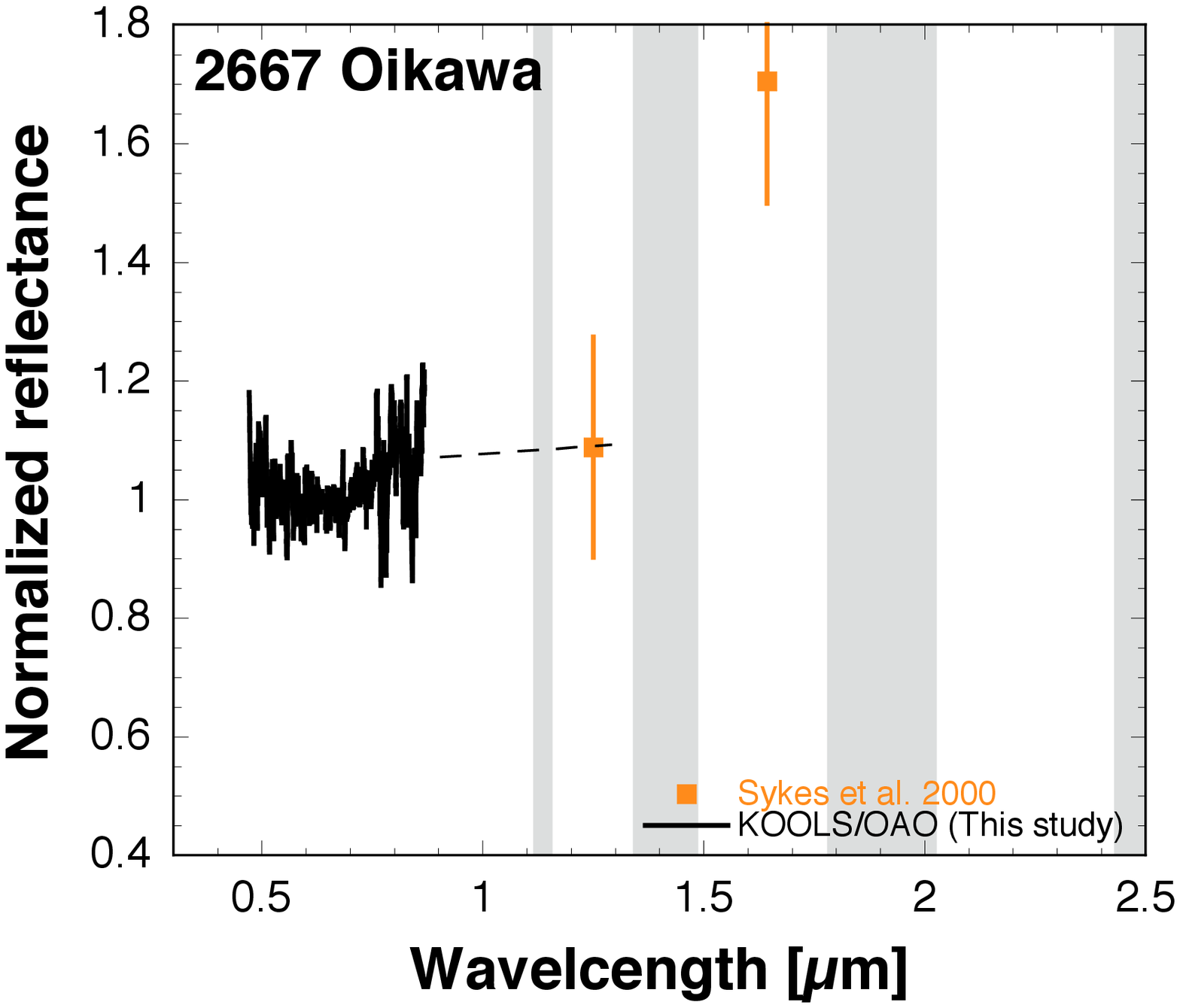}
    \FigureFile(82mm,82mm){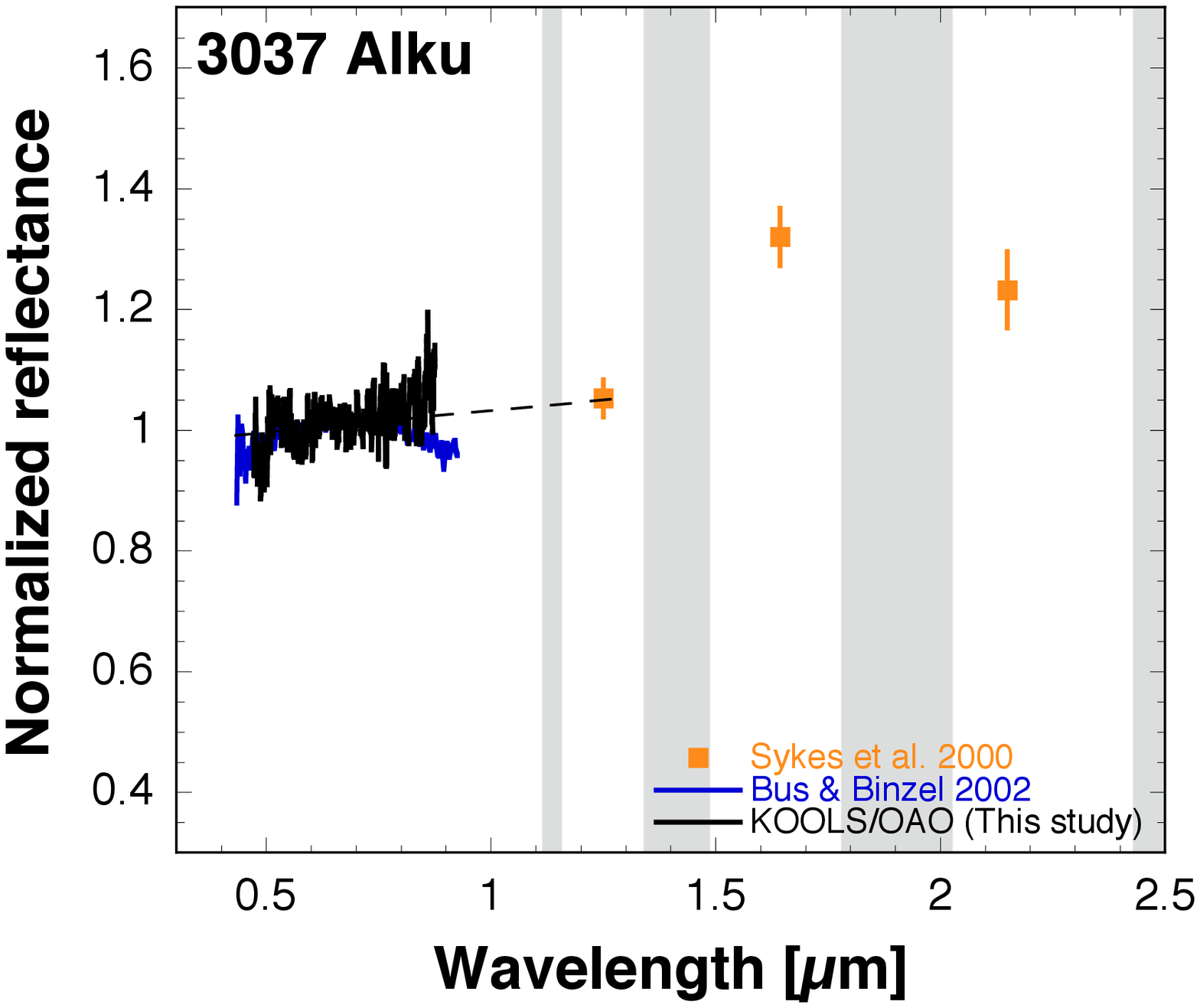}
  \end{center}
  \caption{Continued}
\label{fig:ast3d}
\end{figure*}

The validity of the method in acquiring the visible spectrum of asteroids using KOOLS is shown by comparing the results with the data from the other studies \citep{Hasegawa2014}.
The consistency between the visible spectrum of 3037 Alku in this study (figure \ref{fig:ast3}) and that in \citet{Bus2002a} also supports the validity of the method.
However, the legitimacy of obtaining near-infrared spectral data of asteroids using IRCS with the AO188 system was not disclosed.
To validate the observational methods using IRCS with the AO system, observations for 419 Aurelia, which has a continuous spectrum from 0.3-3.5 \micron\ (\cite{Clark2010}; \cite{Usui2019}), were taken, although it was not in a photometric condition.
The spectra of 419 Aurelia in this study (figure \ref{fig:ast3}) are consistent with those in \citet{Clark2010}, which indicates that obtaining the spectra using IRCS with the AO188 system is the proper method.

The observed asteroids were classified to determine the nature of the bright C-complex asteroids.
To classify the asteroids with only visible data and both visible and near-infrared data, the Bus taxonomy \citep{Bus2002b} and Bus--DeMeo taxonomy (\cite{DeMeo2009}; \cite{Binzel2019}) were employed, respectively.
Based on the Bus taxonomy, nine asteroids were classified by best fitting the mean visible spectra, which are listed on table III in \citet{Bus2002b}, to those data. 
The Bus--DeMeo taxonomy for eighteen asteroids were carried out adapting those data into a taxonomic tool available on the web\footnotemark[5].
When classifying, obvious false data in the spectra due to strong atmospheric absorption between the \textit{Y} and \textit{J} bands, between the \textit{J} and \textit{H} bands, and between the \textit{H} and \textit{K} bands, were corrected by linear interpolation using the appropriate spectra around them.
The Bus--DeMeo classification for five asteroids with visible spectra and near-infrared photometric data in \textit{J}, \textit{H}, and ${K_{S}}$ bands were conducted extrapolating the value in the ${K_{S}}$ band to the value at 2.45 \micron.
Note that both taxonomies are based on the spectral shape only, and geometric albedo is not taken into account for the classification.
\footnotetext[5]{$\langle$http://smass.mit.edu/busdemeoclass.html$\rangle$.}
 
The classification results for bright C-complex, T-type, and unknown spectral type asteroids and for dark (albedo values less than 0.1) asteroids in the Bus and Bus--DeMeo taxonomy are shown in table \ref{tab:tax}.
The results of the Bus classifications for 2250 Stalingrad, 2310 Olshaniya, 2670 Chuvashia, and 4925 Zhoushan could not be narrowed down to one spectral type due to poor quality of the observed data.
By the Bus--DeMeo classification,  Asteroid 555 Norma was classified as ``indeterminate'' using both visible and near-infrared spectra, but was also done as C-type using only near-infrared spectrum.

\begin{landscape}
\setlength{\headsep}{60mm} 
\setlength{\textheight}{160mm} 
\setlength{\tabcolsep}{3pt} 
{
\small
\begin{longtable}{llccccccllll}
  \caption{The Bus and Bus--DeMeo taxonomy of the asteroids in this study.}\label{tab:tax}
  \hline
   No. & Name & \multicolumn{3}{c}{Geometric albedo\footnotemark[$*$]} & Previous & Bus & DeMeo & $\textit{a}$  & $\textit{e}$  & $\textit{i}$  & Family\footnotemark[$\ddagger$]\\
       &      & AKARI & IRAS & WISE & tax\footnotemark[$\dagger$] & tax & tax & [au] & & [$\timeform{D}$]&\\
\endfirsthead
  \hline
  \hline
\endhead
  \hline
\endfoot
  \hline
\multicolumn{1}{@{}l}{\rlap{\parbox[t]{1.0\textwidth}{\small
Top panel: bright C-complex asteroids; second panel: a bright T-type asteroid; third panel: unknown spectral type asteroids; bottom panel: dark asteroids.\\
\footnotemark[$*$]References for geometric albedos: AKARI, \citet{Usui2011}; IRAS, \citet{Tedesco2002a}; WISE, \authorcite{Masiero2011} (\yearcite{Masiero2011}, \yearcite{Masiero2012}).\\
\footnotemark[$\dagger$]References for taxonomy: T84, \citet{Tholen1984}; B02, \citet{Bus2002b}; L04, \citet{Lazzaro2004}; C10, \citet{Carvano2010}.\\
\footnotemark[$\ddagger$]Data for asteroid family are quoted from D. Nesvorn\'y, (2015)\footnotemark[5].\\
\footnotemark[$\S$]The Bus--DeMeo taxonomic result for the asteroid was acquired by extrapolating the value in ${K_{S}}$ band to the value at 2.45 \micron.
}}}
\endlastfoot
  \hline
 723 &Hammonia&{\bf0.294} &{\bf0.183} &{\bf0.352} &C(B02)& --- &{\bf Xn}&2.9935 & 0.0596 & 4.982 &\\
 936 &Kunigunde &{\bf0.124} &{\bf0.113}&0.065 &B(L04)& --- &{\bf C}&3.1302 & 0.1786 & 2.371 &Themis\\
 981 &Martina&{\bf0.108} &{\bf0.125} &0.084 &B(L04)& --- &{\bf B}&3.0950 & 0.2038 & 2.067 &Themis\\
1276 &Ucclia&{\bf0.141} &{\bf0.130} &0.053 &C(L04)& --- &{\bf C}&3.1726 & 0.1021 &23.345 &Alauda\\
1301 &Yvonne&{\bf0.201} &{\bf0.163} &{\bf0.181} &C(B02), Cb(L04), ${C_{\mathrm{p}}}$(C10)\footnotemark[1]& --- &{\bf K}&2.7623 & 0.2734 &34.066 &\\
2519 &Annagerman&{\bf0.105} & ---  &{\bf0.115} &Ch(L04) & --- &{\bf B}&3.1390 & 0.1739 & 2.423 &Themis\\
2525 &O'Steen&{\bf0.124} & ---  &0.098 &B(L04), ${C_{\mathrm{p}}}$(C10)\footnotemark[1]& --- &{\bf Cb}&3.1378 & 0.1906 & 2.774 &Themis\\
2542 &Calpurnia&{\bf0.146} &0.064 &{\bf0.102} &${C_{\mathrm{p}}}$(C10)\footnotemark[1]& --- &{\bf Xk}&3.1296 & 0.0738 & 4.621 &\\
3104 &Durer&{\bf0.237} &--- &{\bf0.357} &Ch(L04), ${A_{\mathrm{p}}}$(C10)\footnotemark[1]& --- &{\bf Sa}&2.9616 & 0.0914 &24.187 &\\
  \hline
979 &Ilsewa&{\bf0.142} &{\bf0.157} &{\bf0.139} &T(L04)& --- &{\bf L}&3.1655 & 0.1339 &10.058 &\\
  \hline
320 &Katharina&{\bf0.165} & ---  &{\bf0.152} &${DL_{\mathrm{p}}}$(C10)\footnotemark[1]&---&{\bf Xe}&3.0096 & 0.1172 & 9.379 & Eos\\
701 &Oriola&{\bf0.239} &{\bf0.218} &{\bf0.124} &  --- & --- &{\bf K}&3.0142 & 0.0336 & 7.143 &\\
840 &Zenobia&{\bf0.610} & ---  &{\bf0.313} & --- & --- &{\bf Sq}&3.1298 & 0.1007 & 9.989 &\\
2250 &Stalingrad&{\bf0.121} & ---  &{\bf0.120} & --- &{\bf B, Cb} &(C)\footnotemark[$\S$]&3.1908 & 0.1801 & 1.524 &Themis\\
2670 &Chuvashia&{\bf0.302} & ---  &{\bf0.246} & --- &X, Xe &{\bf Xe} &3.1688 & 0.0733 & 9.857 &\\
4896 &Tomoegozen&{\bf0.135} &--- &{\bf0.136} & --- &{\bf B} &(C)\footnotemark[$\S$]&3.1070 & 0.1701 &16.565 &\\
4925 &Zhoushan&{\bf0.181} & ---  &{\bf0.195} &${XL_{\mathrm{p}}}$(C10)\footnotemark[1]&{\bf Xe, Xk}&(Xe)\footnotemark[$\S$]&3.0459 & 0.2491 & 8.344 &\\
  \hline
 366 &Vincentina&0.097 &0.080 &0.079 &Ch(B02, L04)& --- &{\bf K}&3.1492 & 0.0554 &10.562 &\\
 419 &Aurelia&0.051 &0.046 & ---  &F(T84), Cb(L04), C(D09)& --- &{\bf C}&2.5958 & 0.2522 & 3.924 &\\
 555 &Norma&{\bf0.101} &0.063 &0.096 &B(B02)& --- &Indet., {\bf C}&3.1844 & 0.1572 & 2.637 &\\
1576 &Fabiola&{\bf0.100} &0.091 &0.075 &BU(T84), B(L04)& --- &{\bf B}&3.1390 & 0.1752 & 0.944 &Themis\\
1764 &Cogshall&0.094 &0.085 &0.061 &  --- &{\bf B}&(C)\footnotemark[$\S$]&3.0945 & 0.1205 & 2.235 &Themis\\
2310 &Olshaniya&0.083 &0.050 &0.064 & --- &{\bf C, Xc} & --- &3.1432 & 0.1613 & 2.650 &Themis\\
2376 &Martynov&0.042 &0.054 &0.034 & --- &{\bf Ch}   & --- &3.2032 & 0.1178 & 3.838 &\\
2667 &Oikawa    &0.041 &0.043 & ---  & --- &{\bf B} & --- &3.2252 & 0.1884 & 2.238 &Themis\\
3037 &Alku&0.061 &0.113 &0.034 &C(B02)&{\bf C}&(C)\footnotemark[$\S$]&2.6754 & 0.1872 &18.980 &\\
\end{longtable}
}
\end{landscape}
\footnotetext[6]{Nesvorn\'y, D. 2015, NASA Planetary Data System, EAR-A-VARGBDET-5-NESVORNYFAM-V3.0}

The DeMeo C-type asteroids include asteroids classified in Bus B-type (see table \ref{tab:HB} in this study; \cite{DeMeo2009}; \cite{deLeon2012}).
In this work, these object are called C-type asteroids with concave curvature.
About one third of the known C-types (DeMeo) asteroids (\cite{DeMeo2009}; \cite{Clark2010}; \cite{Ostrowski2011}; \cite{deLeon2012}; \authorcite{Hasegawa2017} \yearcite{Hasegawa2017}, \yearcite{Hasegawa2018}; \cite{Binzel2019}) are C-types with concave curvature.
It is possible that these data are subject to selection bias, nevertheless, it is worth noting that Bus B-type asteroids are not always classified as DeMeo B-type asteroids. 
This is because the Bus B-type asteroids do not necessarily have concave curvature over the range from the visible to near-infrared.
The Bus B-type asteroids with a flat or convex curvature in the near-infrared are typically classified as DeMeo C-type.

The Bus--DeMeo taxonomic results from both visible and near-infrared spectra for bright C-complex asteroids show that the existence of DeMeo X-complex and C-type with concave curvature, B-, Cb-, K-, and Sa-type (see top of table \ref{tab:tax}).
Bright T-type (Bus) asteroids were classified as L-type (DeMeo) in this study (see second panel of table \ref{tab:tax}).
Bright asteroids with unknown spectrum type were identified as DeMeo X-complex, K-, Sq-, and C-type with concave curvature (see third panel of table \ref{tab:tax}).

\section{The boundaries of albedoes for bright C-complex asteroids}
\begin{figure*}
 \begin{center}
   \FigureFile(82mm,82mm){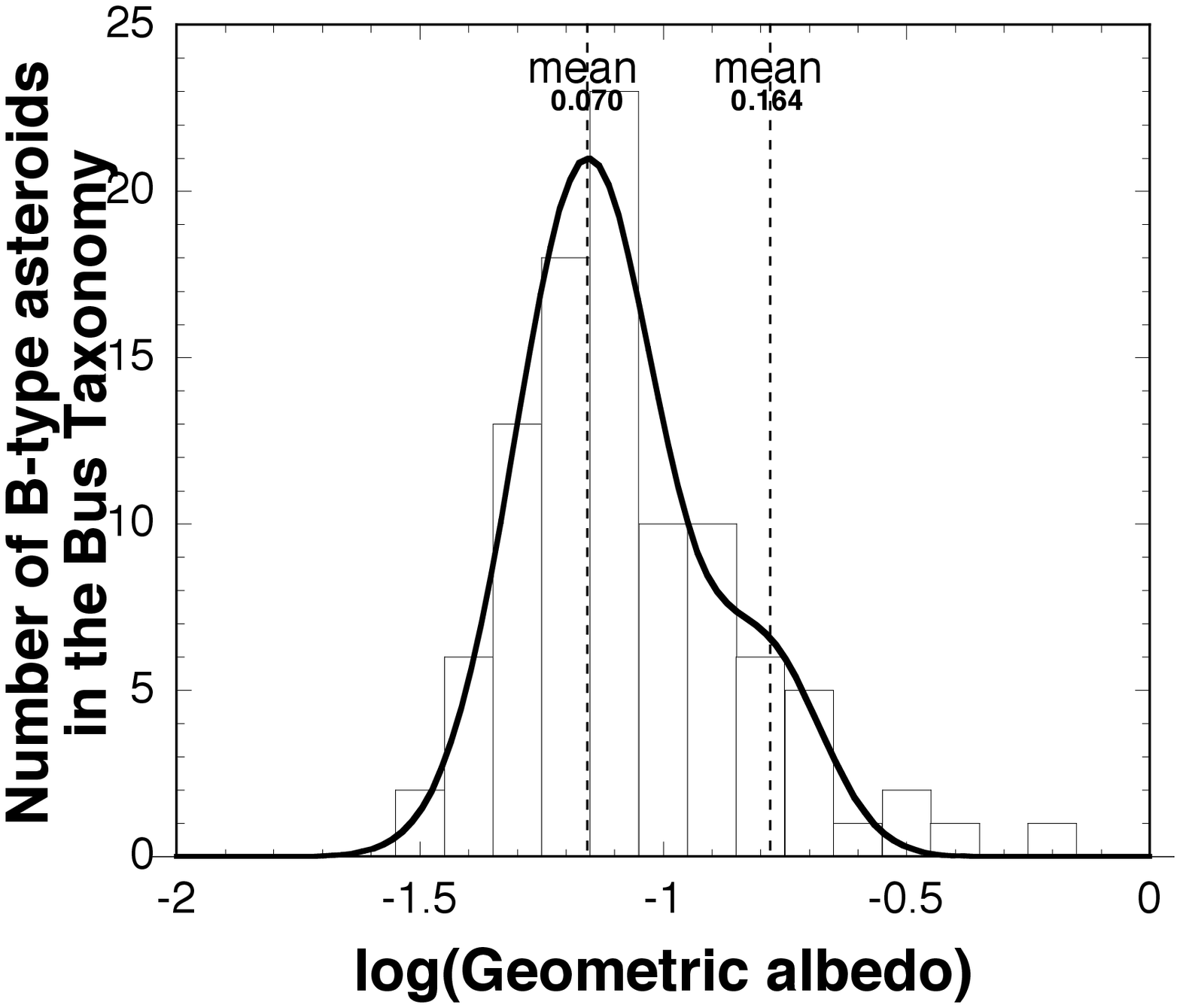}
   \FigureFile(82mm,82mm){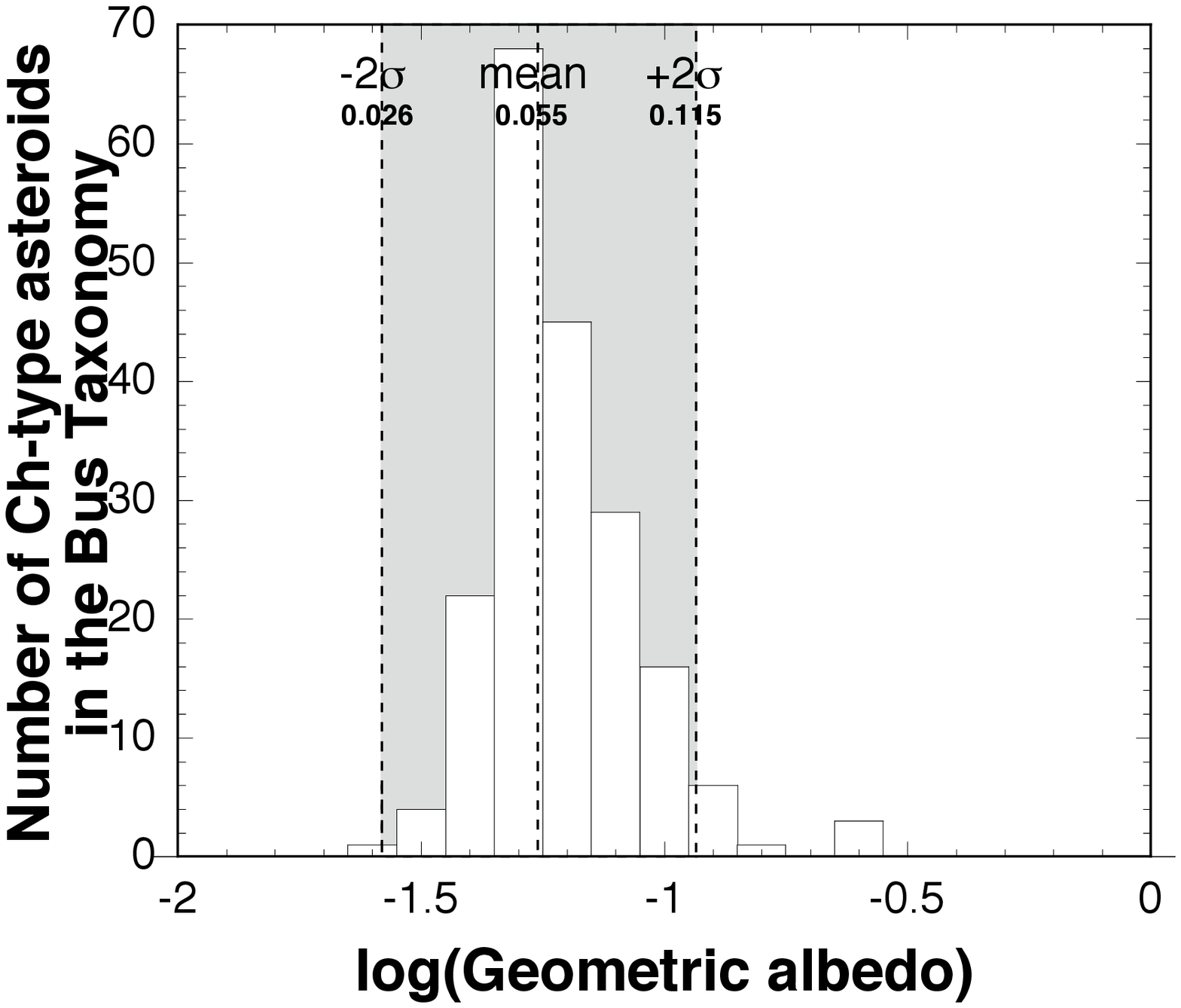}
   \FigureFile(82mm,82mm){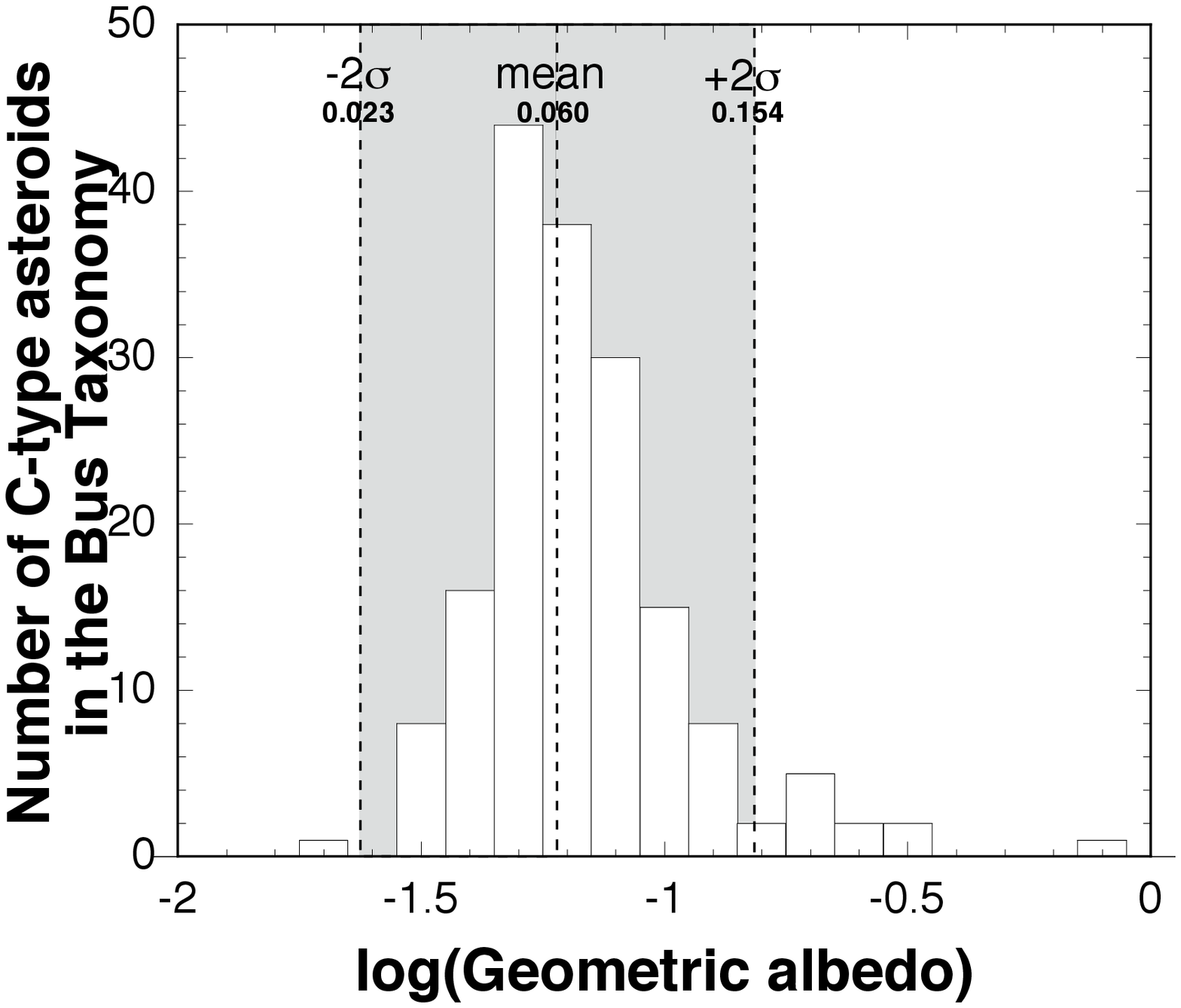}
 \end{center}
  \caption{Albedo distribution of B-, Ch-, and C-type asteroids in the Bus taxonomy. 
The left, middle, and right histograms show the number of B, Ch-, and C-type and asteroids in the Bus taxonomy, respectively.
The number of samples for B-, Ch-, and C-type are 98, 195, and 172, respectively.
The solid line in the figure for B-type asteroids is the best-fitting two-component Gaussian distribution.
}
\label{fig:C1}
\end{figure*}

In order to find appropriate albedo boundaries between the dark and the bright C-complex asteroids, the distributions of the geometric albedo for B-, Ch-, and C-types (Bus) were derived.
The geometric albedo values of asteroids are available in literature (AKARI: \cite{Usui2011}; \cite{Hasegawa2013}; IRAS: \cite{Tedesco2002a}; WISE: \authorcite{Grav2011} \yearcite{Grav2011}, \yearcite{Grav2012}; \cite{Mainzer2011a}; \authorcite{Masiero2011} \yearcite{Masiero2011}, \yearcite{Masiero2012}, \yearcite{Masiero2014}).
The classification results, based on the Bus taxonomy are quoted from \citet{Bus2002b}; \citet{Lazzaro2004}; \authorcite{Binzel2001} \yearcite{Binzel2001}, \yearcite{Binzel2004}; \citet{Rivkin2003}; \citet{Yang2003}; \authorcite{Lazzarin2004a} (\yearcite{Lazzarin2004a}, \yearcite{Lazzarin2004b}, \yearcite{Lazzarin2005}); \citet{Marchi2005}; \citet{Davies2007}; \authorcite{Moskovitz2008a} (\yearcite{Moskovitz2008a}, \yearcite{Moskovitz2008b}); \authorcite{Mothe-Diniz2008a} (\yearcite{Mothe-Diniz2008a}, \yearcite{Mothe-Diniz2008b}); \citet{Roig2008}; \citet{Duffard2009}; \citet{Iwai2017}.
Most asteroids in this study are main-belt asteroids with diameters larger than 10 km.
The albedo distributions are shown in figure \ref{fig:C1}.

In the beginning, we focus on the B-type asteroids.
Figure \ref{fig:C1} (left) indicate the existence of bright B-type asteroids.
\citet{Ali-Lagoa2016} reports that the albedo of the members of the Pallas family occupied by B-type asteroids is 0.141 $\pm$ 0.043.
The presence of bright B-type asteroids (Bus) in the Pallas family  (table \ref{tab:HB}) supports the result of \citet{Ali-Lagoa2016}.
As a result of applying the two-component Gaussian distribution to B-type asteroids, the mean value of the high-albedo part for B-type asteroids is 0.164, which is consistent with the result of \citet{Ali-Lagoa2016}.
The boundary value of the albedo of between the bright and dark B-type asteroids is derived as 0.098.
All B-type asteroids in the Bus taxonomy with albedo values larger than 0.1 are listed in table \ref{tab:HB}. 

Next, the albedo distribution for Bus Ch- and C-type asteroids was statistically evaluated.
In order not to underestimate values of the standard deviation for Ch and C-type (Bus) asteroids, they were obtained without using Gaussian distribution fitting.
The mean and standard deviation of log (albedo) for Ch-type asteroids are $-$1.261 = log (0.065) and 0.160, respectively.
The mean and standard deviation of log (albedo) for C-type asteroids are $-$1.225 = log (0.060) and 0.207, respectively.
Even if the asteroid has an albedo value exceeding 0.1 into the 2$\sigma$ range (Ch-type: less than 0.115, C-type: less than 0.154), it is considered to be a normal Ch- or C-type asteroid.
Figure \ref{fig:C1} (middle and right) shows the results of the albedo distribution for Ch-and C-type (Bus) asteroids.
Almost no Ch- and C-type (Bus) asteroids with an albedo value below $-$2$\sigma$ exist, but there are several Ch- and C-type (Bus) asteroids with albedo values larger than $+$2$\sigma$.
Since Bus Ch- and C-type asteroids with an albedo value within $+$2$\sigma$ can be considered normal Ch- and C-type (Bus) asteroids, we next consider the nature of bright Ch- and C-type (Bus) asteroids with an albedo value exceeding $+$2$\sigma$.
The boundary values of the albedos of between the bright and dark Ch-type asteroids and between the bright and dark C-type asteroids are derived as 0.115 and 0.154
All Ch- and C-type asteroids in the Bus taxonomy with albedo values larger than 0.115 and 0.154 are listed in table \ref{tab:HCh} and \ref{tab:HC}, respectively. 

\begin{longtable}{llllcccl}
  \caption{Bright B-type asteroids in the Bus taxonomy.}\label{tab:HB}
  \hline
   No. & Name & Bus & DeMeo &\multicolumn{3}{c}{Geometric albedo\footnotemark[$\dagger$]} & Remark\\
       &      & Tax\footnotemark[$*$] & Tax\footnotemark[$*$] & AKARI & IRAS, etc & WISE &\\
\endfirsthead
  \hline
  \hline
\endhead
  \hline
\endfoot
  \hline
\multicolumn{1}{@{}l}{\rlap{\parbox[t]{1.0\textwidth}{\small
\footnotemark[$*$]References for taxonomy: B02, \citet{Bus2002b}; L04, \citet{Lazzaro2004}; Ln04b, \citet{Lazzarin2004b}; M08, \citet{Mothe-Diniz2008a}; D09, \citet{DeMeo2009}; Cl10, \citet{Clark2010}; D12, \citet{deLeon2012}; D14, \citet{DeMeo2014}; H17, \citet{Hasegawa2017}; I17, \citet{Iwai2017}.
\\
\footnotemark[$\dagger$]References for geometric albedos: AKARI, \citet{Usui2011}; IRAS, \citet{Tedesco2002a}; WISE, \authorcite{Masiero2011} (\yearcite{Masiero2011}, \yearcite{Masiero2012}, \yearcite{Masiero2019}).\\
\footnotemark[$\ddagger$]Data for asteroid family are quoted from D. Nesvorn\'y, (2015)\footnotemark[6] and data except families were obtained from the NASA JPL HORIZON ephemeris generator system\footnotemark[3].\\
\footnotemark[$\S$]The values are obtained by not IRAS but the NASA Infrared Telescope Facility \citep{Tedesco2005}.\\
\footnotemark[$\|$]Based on the Bus taxonomy, the classifications are re-performed in this study.
}}}
\endlastfoot
  \hline
2&Pallas&B(B02)&B(D09)&{\bf0.159}&{\bf0.143}&{\bf0.142}&Pallas\footnotemark[$\ddagger$]\\
214&Aschera&B(L04), Xc(B02)&Xn(H17)&{\bf0.419}&{\bf0.522}&{\bf0.210}&\\
383&Janina&B(B02)&C(Cl10)&{\bf0.133}&0.093&0.096&Themis\footnotemark[$\ddagger$]\\
400&Ducrosa&B(L04)&---&{\bf0.149}&{\bf0.142}&{\bf0.153}&\\
531&Zerlina&B(B02)&B(d12)&{\bf0.185}&{\bf0.146}&{\bf0.101}&Pallas\footnotemark[$\ddagger$]\\
621&Werdandi&B(L04)&---&{\bf0.124}&{\bf0.153}&{\bf0.127}&Themis\footnotemark[$\ddagger$]\\
936&Kunigunde&B(L04)&C(This study)&{\bf0.124}&{\bf0.113}&0.065&Themis\footnotemark[$\ddagger$]\\
981&Martina&B(L04)&B(This study)&{\bf0.108}&{\bf0.125}&0.084&Themis\footnotemark[$\ddagger$]\\
1003&Lilofee&B(L04)&C(d12)&{\bf0.198}&{\bf(0.212)}\footnotemark[$\S$]&{\bf0.151}&Themis\footnotemark[$\ddagger$]\\
1101&Clematis&B(L04)&---&{\bf0.190}&{\bf0.112}&{\bf0.127}&Alauda\footnotemark[$\ddagger$]\\
1331&Solvejg&B(B02)&---&{\bf0.159}&{\bf0.151}&0.099&\\
1487&Boda&B(L04)&---&{\bf0.133}&{\bf0.120}&0.097&Themis\footnotemark[$\ddagger$]\\
1539&Borrelly&B(B02)&C(d12)&{\bf0.166}&(0.074)\footnotemark[$\S$]&{\bf0.142}&Themis\\
1705&Tapio&B(B02)&C(d12)&{\bf0.100}&{\bf0.118}&0.089&\\
2382&Nonie&B(B02)&---&{\bf0.120}&---&{\bf0.113}&Pallas\footnotemark[$\ddagger$]\\
2464&Nordenskiold&B(L04)&---&0.098&{\bf0.150}&0.083&\\
2525&O'Steen&B(L04)&Cb(This study)&{\bf0.124}&---&0.098&Themis\footnotemark[$\ddagger$]\\
3000&Leonardo&B(B02)&---&---&---&{\bf0.148}&\\
3036&Krat&B(L04)&C(d12)&{\bf0.116}&{\bf0.118}&0.091&Alauda\footnotemark[$\ddagger$]\\
3348&Pokryshkin&B(I17)\footnotemark[$\|$]&---&{\bf0.126}&---&{\bf0.126}&\\
3200&Phaethon&B(B02)&B(d12)&{\bf0.160}&{\bf0.107}&{\bf0.16}&Apollo\footnotemark[$\ddagger$]\\
3579&Rockholt&B(B02)&---&---&---&{\bf0.120}&$\textit{i}$ = 31.1$\timeform{D}$\\
4100&Sumiko&B(L04)&---&{\bf0.156}&---&{\bf0.201}&Eos\footnotemark[$\ddagger$]\\
4396&Gressmann&B(B02)&---&---&---&{\bf0.288}&\\
4896 &Tomoegozen&B(This study) &---&{\bf0.135} &{\bf0.102} &{\bf0.136} &\\
4997&Ksana&B(B02)&---&{\bf0.312}&---&{\bf0.316}&Pallas\footnotemark[$\ddagger$]\\
5222&Ioffe&B(B02)&---&{\bf0.139}&{\bf0.146}&{\bf0.144}&Pallas\footnotemark[$\ddagger$]\\
5234&Sechenov&B(B02)&---&{\bf0.213}&---&{\bf0.157}&Pallas\footnotemark[$\ddagger$]\\
5330&Senrikyu&B(B02)&---&{\bf0.201}&{\bf0.223}&{\bf0.123}&Pallas\footnotemark[$\ddagger$]\\
5639&1989 PE&B(L04)&---&{\bf0.117}&---&{\bf0.198}&$\textit{i}$ = 26.6$\timeform{D}$\\
5870&Baltimore&B(L04)&---&{\bf0.249}&{\bf0.215}&---&Mars-crosser\footnotemark[$\ddagger$]\\
6500&Kodaira&B(B02)&---&---&---&{\bf0.150}&Mars-crosser\footnotemark[$\ddagger$]\\
8567&1996 $\mathrm{HW_{1}}$&B(Ln04b)&Sq(D14)&---&---&{\bf0.160}&Amor\footnotemark[$\ddagger$]\\
9222&Chubey&B(M08)&---&---&---&{\bf0.408}&Tirela\footnotemark[$\ddagger$]\\
49591&1999 $\mathrm{DO_{2}}$&B(I17)\footnotemark[$\|$]&---&---&---&{\bf0.119}&\\
94030&2000 $\mathrm{XC_{40}}$&B(I17)\footnotemark[$\|$]&---&---&---&{\bf0.109}&$\textit{i}$ = 28.0$\timeform{D}$\\
\end{longtable}

\begin{longtable}{llllllll}
  \caption{Bright Ch-type asteroids in the Bus taxonomy.}\label{tab:HCh}
  \hline
   No. & Name & Bus &DeMeo & \multicolumn{3}{c}{Geometric albedo\footnotemark[$\dagger$]} & Remark\footnotemark[$*$]\\
       &      & Tax\footnotemark[$*$] &Tax\footnotemark[$*$] & AKARI & IRAS & WISE &\\
\endfirsthead
  \hline
  \hline
\endhead
  \hline
\endfoot
  \hline
\multicolumn{1}{@{}l}{\rlap{\parbox[t]{1.0\textwidth}{\small
\footnotemark[$*$]References for taxonomy: B02, \citet{Bus2002b}; L04, \citet{Lazzaro2004}; C10, \citet{Carvano2010}.\\
\footnotemark[$\dagger$]References for geometric albedos: AKARI, \citet{Usui2011}; IRAS, \citet{Tedesco2002a}; WISE, \authorcite{Masiero2011} (\yearcite{Masiero2011}, \yearcite{Masiero2012}).\\
\footnotemark[$\ddagger$]Data for asteroid family are quoted from D. Nesvorn\'y, (2015)\footnotemark[6].\\
}}}
\endlastfoot
  \hline
1694&Kaiser&Ch(L04)&---&{\bf0.241}&---&{\bf0.166} &\\
2428&Kamenyar&Ch(B02)&---&{\bf0.139}&0.086&{\bf0.116}&${C_{\mathrm{p}}}$(C10), Veritas\footnotemark[$\ddagger$]\\
3104&Durer&Ch(L04)&Sa(This study)&{\bf0.237}&---&{\bf0.357}&${A_{\mathrm{p}}}$(C10), $\textit{i}$ = 24.2$\timeform{D}$\\
4284&Kaho&Ch(B02)&---&{\bf0.125}&---&{\bf0.128}&${C_{\mathrm{p}}}$(C10)\\
4621&Tambov&Ch(L04)&---&---&---&{\bf0.234}&\\
\end{longtable}

\begin{longtable}{llllllll}
  \caption{Bright C-type asteroids in the Bus taxonomy.}\label{tab:HC}
  \hline
   No. & Name & Bus & DeMeo &\multicolumn{3}{c}{Geometric albedo\footnotemark[$\dagger$]} & Remark\footnotemark[$*$]\\
       &      & Tax\footnotemark[$*$] & Tax\footnotemark[$*$]& AKARI & IRAS, etc & WISE & \\
\endfirsthead
  \hline
  \hline
\endhead
  \hline
\endfoot
  \hline
\multicolumn{4}{@{}l@{}}{\hbox to 0pt{\parbox{85mm}{\footnotesize
\footnotemark[$*$]References for taxonomy: B02, \citet{Bus2002b}; C10, \citet{Carvano2010}; B19, \citet{Binzel2019}.\\
\footnotemark[$\dagger$]References for geometric albedos: AKARI, \citet{Usui2011}; IRAS, \citet{Tedesco2002a}; WISE, \authorcite{Masiero2011} (\yearcite{Masiero2011}, \yearcite{Masiero2012}).\\
\footnotemark[$\ddagger$]Data for asteroid family are quoted from D. Nesvorn\'y, (2015)\footnotemark[6] and data except families were obtained from the NASA JPL HORIZON ephemeris generator system\footnotemark[3].\\
\footnotemark[$\S$]This value was obtained by not IRAS but the NASA Infrared Telescope Facility \citep{Tedesco2005}.
}}}
\endlastfoot
  \hline
723&Hammonia&C(B02)&Xn(This study)&{\bf0.294}&{\bf0.183}&{\bf0.352}&\\
1301&Yvonne&C(B02)&K(This study)&{\bf0.201}&{\bf0.163}&{\bf0.181}&$\textit{i}$ = 34.1$\timeform{D}$\\
1989&Tatry&C(B02)&---&{\bf0.262}&---&{\bf0.193}&\\
2100&Ra-Shalom&C(B04)&B(B19)&{\bf0.177}&---&---&Aten\footnotemark[$\ddagger$]\\
2331&Parvulesco&C(B02)&---&---&---&{\bf0.713}&${CX_{\mathrm{p}}}$(C10)\\
2346&Lilio&C(B02)&---&---&---&{\bf0.277}&\\
2730&Barks&C(B02)&---&{\bf0.196}&{\bf(0.354)}\footnotemark[$\S$]&{\bf0.162}&\\
3137&Horky&C(B02)&---&---&---&{\bf0.207}&${CX_{\mathrm{p}}}$(C10)\\
3435&Boury&C(B02)&---&{\bf0.163}&---&{\bf0.212}&\\
4107&Rufino&C(B02)&---&{\bf0.199}&{\bf0.318}&{\bf0.169}&\\
5678&DuBridge&C(B02)&---&---&---&{\bf0.332}&Pallas\footnotemark[$\ddagger$]\\
\end{longtable}

\section{Discussion}
\subsection{Analogue material and meteorites}
On the basis of visible and near-infrared spectroscopic results in the literature as well as this study, bright C-complex asteroids are classified as B, C with concave curvature, Xn, and Sq (Bus B-type), Sa (Bus Ch-type), and Xn, K and B (Bus C-type) (table \ref{tab:HB}, \ref{tab:HCh}, and \ref{tab:HC}).
There are some outliers in these results. 
For example, Ch-type (Bus) asteroid 3104 Durer is classified into Sa-type in this study, which is consistent with ${A_{\mathrm{p}}}$-type in \citet{Carvano2010}.
Meanwhile, the spectrum of 3104 Durer in Lazzaro et al. (2004) showed Ch-type (Bus) characteristics.
This spectral mismatch may be explained by a misidentification of the target in one of their surveys.
For another example, B-type (Bus) asteroid 8567 1996 $\mathrm{HW_{1}}$ is classified into Sq-type in \citet{DeMeo2014}, which is inconsistent with B-type in \citet{Lazzarin2004b}. 
This spectral discrepancy may also be due to misidentification of the target.
Except for such S-complex asteroids, these spectral type asteroids are candidates for the bright C-complex asteroids. 

It is known that the spectra of meteorites become bluer with an increase in particle size of regolith layer (e.g., \cite{Johnson1973}; \cite{Vernazza2016}).
\citet{Hasegawa2019} argued that the spectra of the Q-type can be explained by lack of fine particles even if space weathering matures their surface.
It is expected that originally bright X-complex asteroids with reddish spectra appear to be bright C-complex asteroids due to spectral bluing effect caused by the absence of regolith on their surface. 
Solar radiation pressure and electrostatic force are known as the effect of removal of fine particles from the asteroid surface layer \citep{Hasegawa2019}.
However, these effects can only be worked effectively to near-earth asteroids with less than 0.5 km in diameter.
Bright C-complex (Bus) asteroids focused in this study are main-belt asteroids with a diameter of more than 10 km, so that fine particles on their surface cannot escape since low influence of these forces affects to the particles due to their high gravity and/or few solar flux compared to that around the near-earth space. 
Therefore, the cause of the existence of bright C-complex (Bus) asteroids cannot be attributed to spectral changes.

Based on meteorite studies (e.g., \authorcite{Hiroi1993} \yearcite{Hiroi1993}, \yearcite{Hiroi1996}), 
bright B-type asteroids are likely to be thermally metamorphosed carbonaceous chondrites. 
\citet{deLeon2012} shows that the spectra of B-type asteroids are similar to those of CV3 and CK4 carbonaceous chondrites.
Meanwhile, \citet{Marsset2020} shows that the cause of the brighter surface on 2 Pallas may be the distribution of salts in the surface layer.
Salts have already been found on 1 Ceres \citep{DeSanctis2016}, and their presence has also been proposed for the Trojan asteroids \citep{Yang2013}.
Xn-type (DeMeo) asteroids are mineralogically associated with enstatite chondrites and/or achondrites (e.g., \cite{Lucas2019}), as also reported in \authorcite{Kasuga2013} (\yearcite{Kasuga2013}, \yearcite{Kasuga2015}).
Enstatite chondrite and/or achondrites are formed in a high-degree reductive environment and a high temperature environment such as in internal igneous or external large-scale impact melting processes (e.g., \authorcite{Keil1989} \yearcite{Keil1989}, \yearcite{Keil2010}).
\citet{Neeley2014} suggested that DeMeo Xk-type asteroids are related to iron meteorites, enstatite chondrites, and stony-iron meteorites, all of which are originated from high-temperature parent bodies. 
DeMeo K-type asteroids are linked to CK chondrites (\cite{Mothe-Diniz2008c}; \cite{Clark2009}), which are known to be highly oxidized and thermally metamorphosed meteorites (e.g., \cite{Greenwood2010}).
As described above, we speculate that the bright C-complex asteroids originated from parent bodies of which the interiors were experienced high temperatures.
To confirm this hypothesis, we investigate the correlation by adapting the results in this study to the known spectrophotometric data as follows. 

\subsection{Sources of the bright C-complex asteroids}

\begin{figure*}
  \begin{center}
    \FigureFile(100mm,100mm){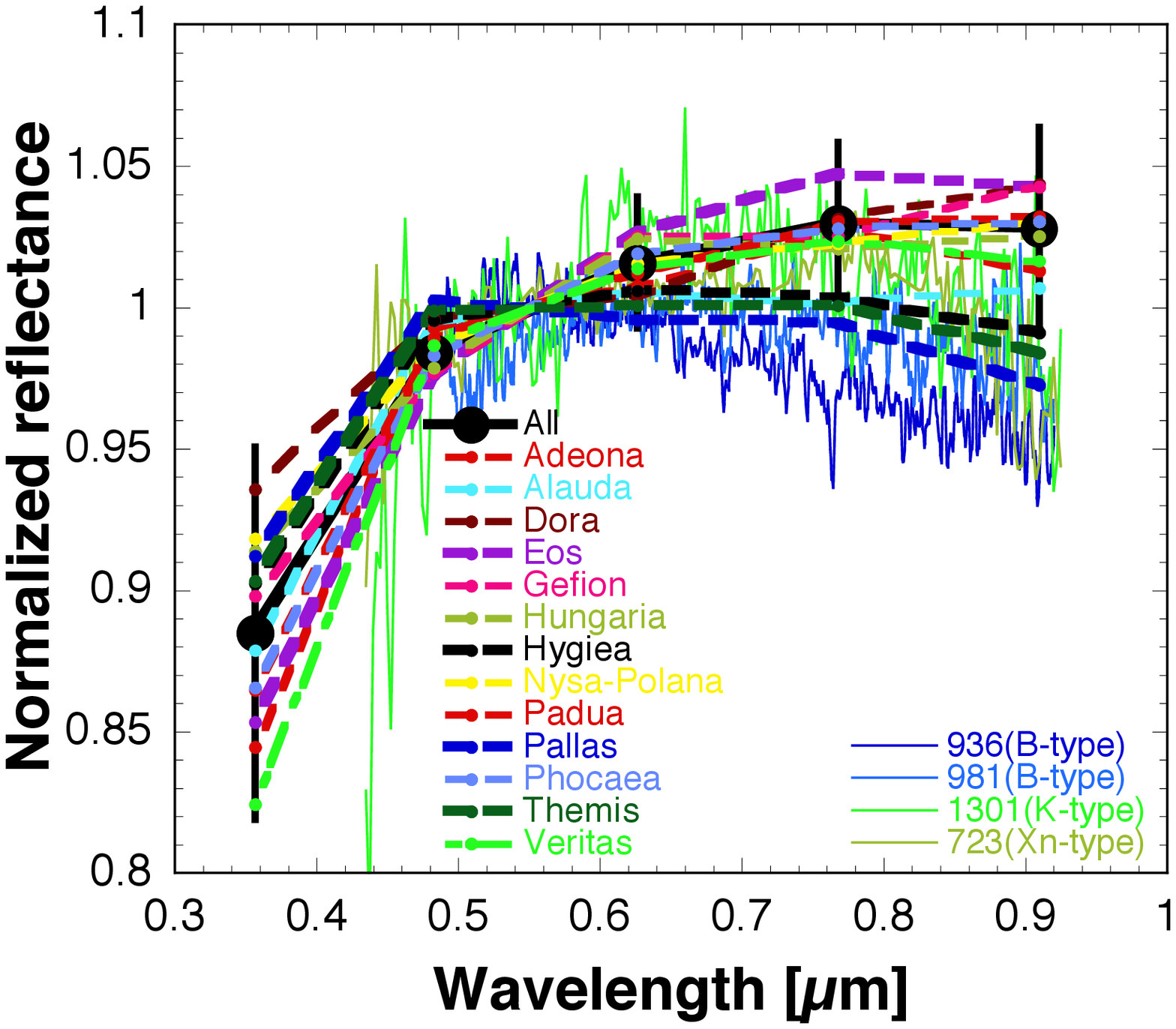}
  \end{center}
  \caption{Spectra for bright C-complex asteroids which are listed in this study and mean spectra of bright ${C_{\mathrm{p}}}$-class asteroids belonging to each family.
}
\label{fig:SDSS}
\end{figure*}

We examine the spectral distribution of some collisional families including bright C-complex asteroids and correlate them with the spectra of the bright C-complex asteroids in order to understand the origin of these asteroids.
Since members of asteroid families are considered as fragments ejected from the parent bodies by catastrophic disruption or cratering, it is possible to estimate the composition of the parent bodies from those spectra. 
The number of spectroscopic data acquired in both the visible and near-infrared wavelength regions for bright C-complex asteroids is still too small and there are not enough numbers of asteroids, even those with visible spectra, to compare the spectroscopic properties of the family members (table \ref{tab:HB}, \ref{tab:HCh}, and \ref{tab:HC}).
To examine the spectral properties of family members, we use the classification based on the Sloan Digital Sky Survey (SDSS) multicolor photometric data in \citet{Carvano2010}. 
The C-complex asteroids in \citet{Carvano2010} are classified into one class, ${C_{\mathrm{p}}}$-type (photometric C-type). 
In this study, we define ${C_{\mathrm{p}}}$-complex asteroids as asteroids of not only ${C_{\mathrm{p}}}$-type, but also ${CD_{\mathrm{p}}}$-, ${CL_{\mathrm{p}}}$-, and ${CX_{\mathrm{p}}}$-types, which are intermediate types between ${C_{\mathrm{p}}}$ and other spectral types.
Bright ${C_{\mathrm{p}}}$-complex asteroids are defined as those with an albedo of more than 0.115 (see section 4), which is the threshold value for bright Ch-type (Bus) asteroids.
We also add the definition of ${B_{\mathrm{p}}}$-type, as B-type asteroids in \citet{Ali-Lagoa2016} among ${C_{\mathrm{p}}}$-complex asteroids.  
The mean spectral distribution and abundance of families with bright C-complex asteroids present in their members  (table \ref{tab:HB}, \ref{tab:HCh}, and \ref{tab:HC}) and families having mean low albedo values are shown in figure \ref{fig:SDSS} and table \ref{tab:SDSST}, respectively.

The Eos family is known to be bright, which have a mean albedo of more than 0.1 \citep{Masiero2011} and a mixture of members with various spectra such as K-, Ld-, Xc-, X-, C-, and B-types \citep{Mothe-Diniz2005}.
Some Eos family members have a flat visible spectrum (e.g., 4455 Ruriko, which is identified as a C-complex asteroid by \cite{Doressoundiram1998}), have the albedo values of more than 0.1 \citep{Masiero2011}. 
In this study, bright ${C_{\mathrm{p}}}$-complex asteroids were found to be abundant in the Eos family (table \ref{tab:SDSST}).
The mean spectrum of bright ${C_{\mathrm{p}}}$-complex asteroids in the Eos family tend to be slightly redder than the spectra of those in other families and also the bright C-complex asteroids in this study (figure \ref{fig:SDSS}).
Within the Eos family, 55 percent of bright ${C_{\mathrm{p}}}$-complex asteroids are of the ${C_{\mathrm{p}}}$-type, 28 percent is of the ${CX_{\mathrm{p}}}$-type and 14 percent is of the ${CL_{\mathrm{p}}}$-type. 
This indicates that the bright ${C_{\mathrm{p}}}$-complex asteroids in the Eos family contains the ${X_{\mathrm{p}}}$-type and the ${L_{\mathrm{p}}}$-type asteroids, and is consistent with the fact that the family members have a diverse spectrum \citep{Mothe-Diniz2005}. 
The Eos family asteroids are thought to be related to the R-chondrites and CK chondrites \citep{Mothe-Diniz2008c}, and these meteorites are known to have formed at high temperature \citep{Weisberg2006}.
Thus, one of the origins of the bright C-complex asteroids is likely to be from asteroids that were formed at high temperature in the interior of the parent bodies.
From the models proposed by \citet{Greenwood2010} for the parent bodies of CV and CK chondrites, it is preferable to adopt one with a single parent object and a single collisional event to explain diverse spectra and the power law of the size distribution of the Eos family members.

Bright ${C_{\mathrm{p}}}$-complex asteroids in the Pallas, Themis, and Hygiea families are found to have concave curvature over 0.5-0.9 microns, similar to the bright B-type asteroids (figure \ref{fig:SDSS}).
Members of these families also include bright ${B_{\mathrm{p}}}$-type asteroids (table \ref{tab:SDSST}).
\citet{Marsset2020} indicated the possibility of salt deposits on 2 Pallas and that on the Pallas family members as well.
The Themis and Hygiea families are located in the outer main-belt, and most of their members have a low albedo of less than 0.1, unlike the Pallas family members (table \ref{tab:SDSST}).
The Hygiea family is thought to have been formed by a cratering event based on the size distribution of its members \citep{Carruba2013}, while the Themis family was formed by a catastrophic disruption event \citep{Marzari1995}, both of which were originated from the parent bodies of similar size to that of 2 Pallas \citep{Vernazza2020,Marzari1995}.
The composition of these parent bodies could be CK and/or R chondrite analogues, such as the Eos family, or salt deposits, such as the 2 Pallas \citep{Marsset2020}.
Besides, (a) salts may be present in asteroids in the outer main-belt \citep{Poch2020}, (b) the theory and observations suggest that salts may be present in the Themis parent body (\cite{Castillo-Rogez2010}; \cite{Marsset2016}), (c) salt precipitation can be explained by small percentages \citep{Castillo-Rogez2018}, thus it is very likely that the parent bodies of the Themis and Hygiea family were as likely to have precipitated salt as 1 Ceres and 2 Pallas. However, the proportion of salts produced in Themis and Hygiea parent bodies is considered to be lower than that in 2 Pallas, because a fraction of bright C-complex asteroids in members of the Themis and Hygiea family are smaller than those of the Pallas family (table \ref{tab:SDSST}).

The spectral trend of bright ${C_{\mathrm{p}}}$ asteroids in other families shown in table~\ref{tab:SDSST} is similar to the spectral trend for all bright ${C_{\mathrm{p}}}$ asteroids. 
This might be caused by common process, not by common composition;  
one of the candidates of the common process is thermal metamorphism due to frequently impact events. 
Carbonaceous chondrites are known to get brighter albedo values upon thermal metamorphism (\cite{Hiroi1993}; \cite{Cloutis2012}).
The area where the impact event raises the temperature to occur such a metamorphism is limited to several times as large as the size of the projectile at most (e.g., \cite{Davison2014}; \cite{Kurosawa2018}). In addition, the area around the impact point becomes fine ejecta, impact metamorphosed asteroids may not remain as large objects. 
Thus the ratio of bright ${C_{\mathrm{p}}}$-class asteroids in the family composed of low albedo asteroid members is considered to be small as a few percent (table \ref{tab:SDSST}).

The ratio of all bright ${C_{\mathrm{p}}}$-complex asteroids in the total population is different from that in low albedo families. 
This might be caused by other factors than impact thermal metamorphism.
Since DeMeo K-type and Xn-type asteroids were identified as spectral type of bright C-complex asteroids in this study, bright ${C_{\mathrm{p}}}$-complex asteroids not belonging to any families are considered to be composed of CV/CK chondrites, enstatite chondrites/achondrites, or impact metamorphosed carbonaceous chondrites.

\begin{longtable}{lcccccl}
  \caption{Proportion of bright C-complex asteroids in each family based on SDSS spectroscopic data in \citet{Carvano2010}.}\label{tab:SDSST}
  \hline
 Family&All &Bright &Ratio of bright&Albedo&Typical&Region\\
 &${C_{\mathrm{p}}}$-comp\footnotemark[$*$]&${C_{\mathrm{p}}}$-comp\footnotemark[$*$]&${C_{\mathrm{p}}}$-comp&distribution\footnotemark[$\dagger$]&spectral type\footnotemark[$\ddagger$]&\\
\endfirsthead
  \hline
  \hline
\endhead
  \hline
\endfoot
  \hline
\multicolumn{1}{@{}l}{\rlap{\parbox[t]{1.0\textwidth}{\small
\footnotemark[$*$]Numbers in parentheses are the number of ${B_{\mathrm{p}}}$-types in \citet{Ali-Lagoa2016} for which data are listed in \citet{Carvano2010}.\\
\footnotemark[$\dagger$]References for albedo disrtribution: Alauda family, \citet{Masiero2015}; Phocaea family, \citet{Novakovic2017}; other, \citet{Masiero2011}.\\ 
\footnotemark[$\ddagger$]References for typical spectral type: Alauda family, \citet{Hsieh2018}; Gefion family, \citet{Mothe-Diniz2005}; Nysa-Polana family, \citet{Cellino2001}; Tirela family, \citet{Mothe-Diniz2008b}; other, \citet{Masiero2015}.\\
}}}
\endlastfoot
  \hline
All        &11714(780)&1509(25)&0.129 &&\\
  \hline
Adeona     &226(17)   &4(0)    &0.018 &Low&Ch&Central main-belt\\
Dora       &163(10)   &3(0)    &0.018 &Low&Ch&Central main-belt\\
Hoffmeister&140(5)    &0(0)    &0.000 &Low&C&Central main-belt\\
Padua      &85(4)     &3(1)    &0.035 &Low&X&Central main-belt\\
Hygiea     &449(86)   &27(2)   &0.060 &Low&C&Outer main-belt\\
Themis     &550(160)  &21(3)   &0.038 &Low&C&Outer main-belt\\
Veritas    &175(1)    &6(1)    &0.034 &Low&Ch&Outer main-belt\\
Alauda     &199(25)   &7(0)    &0.035 &Low&B,C,X&Outer main-belt, large \it{i}\\
  \hline
Eos        &241(0)    &156(0)  &0.647 &Middle&K&Outer main-belt\\
Pallas     &15(7)     &10(6)   &0.667 &Middle&B&Central main-belt, large \it{i}\\
  \hline
Nysa-Polana&283(41)   &20(0)   &0.071 &Bimodal&F,S&Inner main-belt\\
Gefion     &17(0)     &2(0)    &0.118 &Bimodal&S,C&Central main-belt\\
Tirela     &6(0)      &0(0)    &0.000 &Bimodal&L,Ld&Outer main-belt\\
Phocaea    &18(0)     &4(0)    &0.222 &Bimodal&C,S&Inner main-belt, large $\it{i}$\\
  \hline
Hungaria   &5(0)      &5(0)    &1.000 &High&Xe&Inside of main-belt, large $\it{i}$\\
\end{longtable}

\section{Conclusions}
Spectroscopic observations of bright C-complex asteroids were carried out to unravel the origin and properties of asteroids.
%
The results suggest that most bright C-complex (Bus) asteroids are composed of DeMeo C-type with concave curvature, B-type, Xn-type and K-type asteroids.
The meteorites associated with these spectral asteroids have been found to be composed of minerals that have experienced high temperatures.
Based on the comparison of observations in this study with the spectra of families, the origin of the bright C-complex asteroids can be inferred as follows.
\begin{description}
\item{$\bullet$}
The member of the Eos family, including the bright C-complex asteroids, are composed of CV/CK chondrite analogues.
\item{$\bullet$}
Salts may have been deposited in the Pallas, Themis and Hygiea parent bodies.
It is likely that there was not much salts present in the parent body of the Themis and Hygiea family due to the paucity of bright C-complex asteroids in the families.
On the contrary, it is possible that 2 Pallas had more salts than the Themis and Hygiea parent bodies since the Pallas family composes more bright C-complex asteroids.
\item{$\bullet$}
For other families occupied by dark C-complex asteroids, the bright C-complex asteroids were most likely formed by impact thermal metamorphism.
\item{$\bullet$}
Bright C-complex asteroids that do not belong to any families could be linked with CV/CK chondrites and/or enstatite chondrites/achondrites, besides impact metamorphosed carbonaceous chondrites.
\end{description}


\bigskip
\begin{ack}
We are grateful to Dr. V\'ictor M. Al\'i-Lagoa and Dr. Marcela I. Ca\~nada Assandri for sharing valuable theirs SDSS photometric data. 
We would like to thank the referee for her/his careful reviewing and giving constructive suggestions, which helped us to improve the manuscript significantly.
We are grateful to Dr. Masateru Ishiguro and Dr. Katsuhito Ohtsuka for insights and fruitful discussions. 
This study was based in part on data collected at Okayama Astrophysical Observatory and using the Subaru Telescope, which is operated by NAOJ. 
Data collected at Okayama Astrophysical Observatory are used, as well as data obtained from SMOKA, which is operated by the Astronomy Data Center, NAOJ.
Taxonomic type results presented in this work were determined, in whole or in part, using a Bus--DeMeo Taxonomy Classification Web tool by Dr. Stephen. M. Slivan, developed at MIT with the support of National Science Foundation Grant 0506716 and NASA Grant NAG5-12355.
This study has utilized the SIMBAD database, operated at CDS, Strasbourg, France and the JPL Small-Body Database Browser, operated at JPL, Pasadena, USA.
We would like to express our gratitude to the staff members of Okayama Astrophysical Observatory and the Subaru Telescope for their assistance.
DK is supported by Optical \& Near-Infrared Astronomy Inter-University Cooperation Program, the MEXT of Japan.
This study was supported by JSPS KAKENHI (grant nos. JP15K05277, JP17K05636, JP18K03723, JP19H00719, JP19H00725, JP20H00188, and JP20K04055) and by the Hypervelocity Impact Facility (former facility name: the Space Plasma Laboratory), ISAS, JAXA.
\end{ack}


\end{document}